\documentclass[journal]{IEEEtran}

\IEEEoverridecommandlockouts

\usepackage[utf8]{inputenc}
\usepackage[T1]{fontenc}
\usepackage[american]{babel}
\usepackage{graphicx}
\usepackage{listings}
\usepackage{float}
\usepackage{amsmath,amssymb,exscale}
\usepackage{blindtext, graphicx}
\usepackage{verbatim}
\usepackage{algorithm}
\usepackage{algpseudocode}

\usepackage{fancyvrb}
\usepackage{bera}
\usepackage{amsmath}
\usepackage{mathtools}
\usepackage{lipsum}

\usepackage{times}
\usepackage{scalerel}
\usepackage{tikz}
\usetikzlibrary{svg.path}
\usepackage[bookmarks=false]{hyperref}

\usepackage{booktabs}
\usepackage{multirow}
\usepackage{resizegather}

\usepackage{soul}
\usepackage{subcaption}
\usepackage{setspace}

\usepackage{caption}
\graphicspath{{Images/}}
\begin{document}

\author{
       \IEEEauthorblockN{
       Xiaoqi Zhang,
       Zhitong Ni, \IEEEmembership{Member, IEEE},
       Weijie Yuan, \IEEEmembership{Senior Member, IEEE}, \\ 
       J. Andrew Zhang, \IEEEmembership{Senior Member, IEEE}, and Tony Q. S. Quek, \IEEEmembership{Fellow, IEEE}
       }
   \thanks{This work was supported in part by the Australian Government through the Australian Research Council's Discovery Projects funding scheme (project DP210101411); in part by the National Research Foundation, Singapore and Infocomm Media Development Authority under its Communications and Connectivity Bridging Funding Initiative.}
   \thanks{X. Zhang and J. Andrew Zhang are with the School of Electrical and Data Engineering, University of Technology Sydney, Sydney, 2007, Australia (e-mail: Xiaoqi.Zhang@student.uts.edu.au, andrew.zhang@uts.edu.au).}
   \thanks{Z. Ni is with the School of Electrical Engineering and Telecommunications, University of New South Wales, Sydney, NSW 2052, Australia (e-mail: zhitong.ni@unsw.edu.au).}
   \thanks{W. Yuan is with the School of Automation and Intelligent Manufacturing, Southern University of Science and Technology, Shenzhen 518055, China. (e-mail: yuanwj@sustech.edu.cn).}
   \thanks{T. Q. S. Quek is with the Singapore University of Technology and Design, Singapore 487372, and also with the Department of Electronic Engineering, Kyung Hee University, Yongin 17104, South Korea (e-mail: tonyquek@sutd.edu.sg).}
   \vspace{-1.0cm}
   
}

\title{Deep Learning-based OTFS Channel Estimation and Symbol Detection with Plug-and-Play Framework}

\maketitle

\begin{abstract}
    Orthogonal Time Frequency Space (OTFS) modulation has recently attracted significant interest due to its potential for enabling reliable communication in high-mobility environments. However, the effectiveness of OTFS receivers relies on the inherent characteristic of the Delay-Doppler (DD) domain channel, where the sparsity of the discretized channel varies across different communication scenarios. For instance, the fractional Doppler effect reduces the inherent channel sparsity, which consequently degrades channel estimation accuracy and increases the complexity of symbol detection. Traditional algorithms relying on fixed sparsity priors often require manual design, while purely data-driven deep learning (DL) methods typically struggle to generalize across diverse channel conditions. To address these challenges, we propose a novel unsupervised DL-based plug-and-play (PnP) framework that provides a flexible solution for OTFS receiver design. The proposed framework can be applied to both channel estimation and symbol detection, jointly leveraging the flexibility of optimization-based methods and the powerful generalization capability of data-driven models. Specifically, a lightweight encoder-decoder network (EDN) is incorporated as an implicit channel prior for channel estimation, enabling robust performance across varying levels of channel sparsity. Furthermore, for symbol detection, we realize the PnP framework with a time-domain matrix inversion for model-based equalization, followed by a small multi-layer perceptron (MLP) pre-trained for specific constellations, thereby achieving low complexity and enabling flexible adaptation to various modulation formats. Finally, numerical results demonstrate the effectiveness and robustness of the algorithm.
 
\end{abstract}

\vspace{-0.2cm}

\section{Introduction}

The forthcoming beyond 5G (B5G) wireless communication systems are envisioned to support a range of emerging high-mobility communication applications, such as low Earth orbit (LEO) satellites, unmanned aerial vehicles (UAVs), and autonomous cars \cite{yuan2020simple,ni2021uplink}. The conventional long-term evolution (LTE) modulations, such as orthogonal frequency-division multiplexing (OFDM), demonstrate limited capability in maintaining efficient and reliable communication performance under high-mobility channel conditions. This limitation primarily stems from the increased frequency dispersion induced by Doppler shift effects, which significantly degrades the system performance in such scenarios. To accommodate the heterogeneous demands of B5G wireless systems in high-mobility scenarios, researchers have proposed new modulation techniques and waveforms, e.g., orthogonal chirp
division multiplexing (OCDM), affine frequency division multiplexing (AFDM), and orthogonal time frequency space (OTFS) modulations \cite{wei2021waveform}. {Amongst them, OTFS is specifically designed to cope with high-mobility channels, where Doppler effects become dominant, and has therefore attracted significant attention as a promising solution for ultra-reliable communications.} {While AFDM also exhibits strong potential for supporting high-speed and robust transmission, OTFS provides a well-established delay–Doppler (DD) domain framework that has been extensively adopted for channel modeling and receiver design. Hence, this work focuses on OTFS, and the extension to other waveforms is left for future research.}\par  

In high-mobility scenarios, wireless channels undergo dual dispersion effects in the time-frequency (TF) domain, where time dispersion arises from multipath propagation, while frequency dispersion is induced by Doppler shifts. Traditional OFDM modulation effectively mitigates inter-symbol interference (ISI) through cyclic prefix (CP) implementation. However, the inherent subcarrier orthogonality in OFDM systems is severely degraded by inter-carrier interference (ICI), ultimately leading to unsatisfactory reliability. In contrast to OFDM, OTFS modulation, grounded in Zak transform theory, operates in the DD domain, exhibiting quasi-static, sparse, and compact channel characteristics \cite{wei2021waveform}. Furthermore, each data symbol in the DD domain undergoes the entire channel fluctuation in the TF domain over an OTFS. The unique characteristic facilitates the full exploitation of channel diversity, which is essential for achieving the ultra-high reliability required by B5G wireless communication systems. \par 

Despite the significant potential of OTFS in supporting reliable communication over high-mobility channels, practical receiver design still faces major challenges in both channel estimation and symbol detection \cite{wang2022joint, nie2024uplink}. A key challenge is the need for accurate channel estimation, which is essential for effective equalization and reliable signal detection. Traditional estimation methods, such as least squares (LS) and linear minimum mean square error (LMMSE), fail to fully exploit the inherent sparsity of the DD domain channel, driving the development of sparsity-aware channel estimation methods. For instance, \cite{raviteja2019embedded} introduced an embedded pilot-aided channel estimation framework that leverages a single pilot impulse with guard zero symbols in the DD domain, capitalizing on channel sparsity to improve performance. Similarly, off-grid sparse Bayesian learning (SBL) has been employed to mitigate channel spreading caused by fractional delay and Doppler shifts, thereby enhancing estimation accuracy by focusing on the original DD domain channel response \cite{wei2022offgrid}. Moreover, 3D-structured orthogonal matching pursuit (OMP) algorithms have been developed to effectively exploit the inherent multi-dimensional sparsity of the OTFS massive MIMO channel across the delay, Doppler, and angle dimensions for efficient downlink channel estimation \cite{shen2019channel}. However, a practical challenge arises because the DD channel sparsity varies with OTFS parameters and propagation conditions, particularly due to fractional Doppler effects, which disrupt the sparsity structure and degrade estimation accuracy \cite{wei2021transmitter}. Furthermore, while the widely used message-passing algorithm (MPA) for symbol detection has shown promise \cite{li2023otfs, zhongjie2022uamp}, it suffers from high computational costs as sparsity decreases. Additionally, traditional detection methods remain highly sensitive to channel estimation errors, further impacting their performance \cite{ye2017power}. These limitations highlight the need for a flexible channel estimation method and a low-complexity symbol detection mechanism capable of adapting to dynamic changes in channel sparsity and functioning effectively with potentially imperfect channel state information.

In the domain of wireless communications, deep learning (DL) has emerged as a transformative approach, demonstrating data-driven capabilities across various applications, including channel estimation, symbol detection, and spectrum sensing \cite{Liu2022RIS}. Specifically, a novel deep residual network was proposed in \cite{xiaoqi2024sparse} for OTFS channel estimation. This approach exploits the inherent sparsity characteristics of the channel, demonstrating enhanced estimation accuracy, particularly in non-ideal transmission scenarios. Moreover, \cite{qing2023viterbi} developed a Viterbi-based neural network (ViterbiNet) for OTFS symbol detection, which achieves remarkable performance with minimal network size and training data requirements. Nonetheless, existing methods still struggle with limited adaptability to diverse and dynamic wireless environments, often requiring task-specific models that lack generalization. In addition, purely data-driven approaches often lack interpretability and necessitate a large number of parameters to learn task-specific mappings, which increases their inefficiency and complexity. \par
{Fortunately, the plug-and-play (PnP) prior framework has become a widely applied approach in imaging that couples a fixed data-fidelity operator with a replaceable denoiser \cite{zhang2021plug}. This modularity supports transfer across tasks by lightly adapting the denoiser while keeping the physical consistency step unchanged. At the same time, the denoiser serves as an expressive implicit prior, where encoder–decoder and recurrent architectures capture clustered and multi-scale structures and are trained over ranges of noise strengths, which improves robustness under distribution shifts and imperfect measurements \cite{lai2022deep}. Empirical studies demonstrate strong performance in deblurring, super-resolution, compressed sensing, and inpainting, and theoretical analyses further examine stability and convergence under mild conditions \cite{sreehari2016plug,ryu2019plug}. These properties align with the need for flexible algorithms that retain a stable data-consistency core while adapting the prior to changing statistics.}\par


In this paper, we propose a unified DL-based PnP framework to address both channel estimation and symbol detection in OTFS systems. By integrating the physical consistency of model-based operators with a DL-based denoiser, our approach provides a robust and efficient receiver architecture capable of handling diverse and challenging channel conditions. The main contributions of this paper are summarized as follows: {
\begin{itemize}
    \item Unlike conventional methods that rely on distinct, task-specific solutions, such as the manually designed priors in traditional algorithms or the monolithic networks in end-to-end DL, we formulate both tasks as linear inverse problems and introduce a unified deep PnP framework. Specifically, this framework employs variable splitting to decompose the problem into a model-based subproblem that enforces data fidelity according to the system model, and a learning-based denoising module that serves as an adaptive prior. This unified design provides a systematic way to integrate the physical consistency of model-driven operators with the flexibility of learning-based priors, thereby enhancing the adaptability across diverse channel conditions.
    \item Building upon the unified PnP framework, we propose a PnP-based channel estimation (PnP-CE) method tailored for the OTFS channel. Motivated by the limitations of traditional estimators that rely on fixed sparsity assumptions, we realize the learning-based denoiser using a lightweight encoder–decoder network (EDN) that adaptively exploits the sparsity and leakage characteristics of DD channels. By learning the intricate channel structure, the proposed PnP-CE method significantly improves estimation accuracy and maintains robustness across diverse channel conditions without requiring retraining. 
    \item For the symbol detection task, we implement the unified PnP framework as a PnP-based symbol detector (PnP-SD) featuring a hybrid-domain architecture. The core of this detector consists of a model-based equalization step performed via an efficient matrix inversion in the time domain, and a learning-based denoising module realized by a constellation-aware multi-layer perceptron (MLP). The time-domain inversion leverages the block-diagonal structure to substantially reduce computational complexity, while the specialized MLP denoiser utilizes constellation priors to enhance robustness against imperfect channel state information (CSI), resulting in a detector that is both efficient and accurate for practical systems.
    \item Extensive simulations are conducted to evaluate the proposed framework for both channel estimation and symbol detection. The results highlight the scalability of the PnP framework, demonstrating robust channel estimation performance across integer and fractional Doppler cases, while also showing that PnP-SD achieves improved detection accuracy and reduced complexity under non-ideal channel conditions.
\end{itemize}}

The remainder of this work is organized as follows. Section \uppercase\expandafter{\romannumeral2} presents the system model for OTFS modulation, along with the input-output relationship with bi-orthogonal and rectangular waveforms. Section \uppercase\expandafter{\romannumeral3} formulates the problems of channel estimation and symbol detection. Section \uppercase\expandafter{\romannumeral4} presents the design of a unified deep PnP framework for the OTFS receiver. As an implementation of the proposed framework, Section \uppercase\expandafter{\romannumeral5} introduces a PnP-CE algorithm and a PnP-SD algorithm. To evaluate the effectiveness of the proposed approach, extensive simulation results are provided in Section \uppercase\expandafter{\romannumeral6}. Finally, the key findings of this work are summarized in \uppercase\expandafter{\romannumeral7}. \par 

\textit{Notations:} The following notations are used in this work. Scalars, vectors, and matrices are denoted by regular font (i.e., \( x \)), bold lowercase letters (i.e., \( \mathbf{x} \)), and bold uppercase letters (i.e., \( \mathbf{X} \)), respectively. The sets \( \mathbb{C}^{M \times N} \) and \( \mathbb{R}^{M \times N} \) represent complex-valued matrices and real-valued matrices of size \( M \times N \), respectively. The symbols \( \Re(\cdot) \) and \( \Im(\cdot) \) represent the real and imaginary components of an input complex number. The real-valued Gaussian distribution and the circularly symmetric complex Gaussian (CSCG) distribution are represented by \( \mathcal{N}(\boldsymbol{\mu}, \boldsymbol{\Sigma}) \) and \( \mathcal{C}\mathcal{N}(\boldsymbol{\mu}, \boldsymbol{\Sigma}) \), where \( \boldsymbol{\mu} \) and \( \boldsymbol{\Sigma} \) denote the mean vector and the covariance matrix, respectively. The notation \( \mathbf{I}_N \) is the identity matrix of size \( N \times N \), and \( \mathbf{1}_N \) is the column vector of all ones with size \( N \times 1 \). \( (\cdot)^{T} \), \( (\cdot)^{H} \), and \( (\cdot)^{*} \) stand for transpose, Hermitian transpose, and conjugate operations, respectively. The notation \( \mathcal{O}(\cdot) \) represents the order of computational complexity. 
\section{System Model} 

\begin{figure}[tb]
    \centering
    \includegraphics[width=0.8\linewidth]{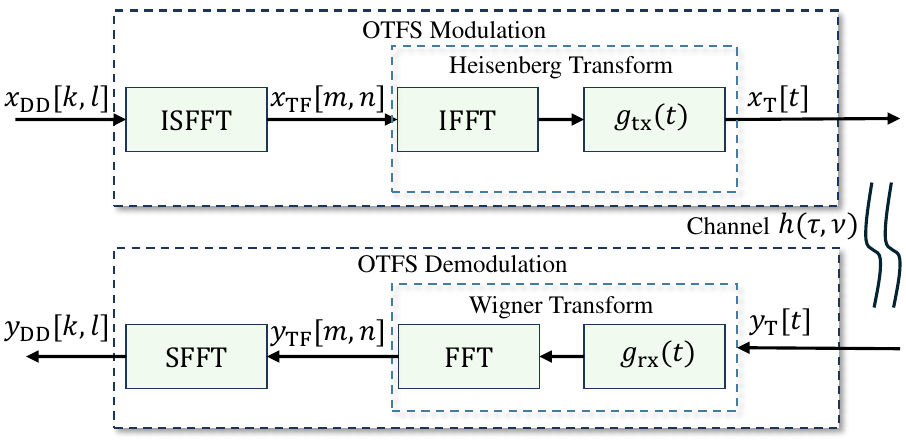}
    \caption{Block diagram of the considered OTFS system.}
    \label{OTFS_M}
\end{figure}

\subsection{OTFS Modulation}
We consider an OTFS modulation as shown in Fig. \ref{OTFS_M}. Let $N$ denote the number of time slots and $M$ the number of subcarriers for each OTFS symbol. The data symbols $x_{\mathrm{DD}}[k,l]$, with Doppler index $k\in\{0, \cdots, N-1\}$ and delay index $l \in \{0, \cdots, M-1\}$, are selected from a constellation set $\mathbb{Q}=\{x_1,\cdots,x_{Q'}\}$ of size $Q'$ and are placed in the DD domain. The TF domain transmitted symbol $x_{\mathrm{TF}}[m,n]$, with time index $n \in \{0, \dots, N-1\}$ and frequency index $m \in \{0, \dots, M-1\}$, is obtained through the inverse symplectic finite Fourier transform (ISFFT), i.e., 
\begin{align}
    x_\mathrm{TF}\left[m,n\right]=\frac1{\sqrt{NM}}\sum_{k=0}^{N-1}\sum_{l=0}^{M-1}x_\mathrm{DD}\left[k,l\right]e^{j2\pi\left(\frac{nk}N-\frac{ml}M\right)}.
\end{align}

The time domain transmitted OTFS signal $x_\mathrm{T}(t)$ is derived from
\begin{align}
    x_\mathrm{T}(t)=\sum_{n=0}^{N-1}\sum_{m=0}^{M-1}x_{\mathrm{TF}}[m,n]g_{\mathrm{tx}}(t-nT)e^{j2\pi m \Delta f(t-nT)},
\end{align}
where $T$ and $\Delta f$ are the TF domain time slot duration and subcarrier spacing, respectively. The notation $g_{\mathrm{tx}}(t)$ denotes the pulse shaping filter. We consider the OTFS signal transmitting over a time-varying channel, whose impulse response can be characterized with a DD domain representation as 
\begin{align}
    h(\tau, \nu) = \sum_{i=1}^P h_{i} \delta(\tau-\tau_{i})\delta(\nu-\nu_{i}),
    \label{ideal_channel}
\end{align}
where $P$ is the number of paths, and $h_i$, $\tau_i$ and $\nu_i$ denote the channel coefficients, delay, and Doppler shifts corresponding to the $i$-th path, respectively. Specifically, the delay shifts $\tau_i$ and Doppler shifts $\nu_i$ associated with the $i$-th path can be expressed as $\tau_i=\frac{l_i}{M \Delta f}$,  $\nu_i=\frac{k_i+\kappa_i}{NT}$. Here, integers $l_i \in [0, l_{\max}]$ and $k_i\in [-k_{\max}, k_{\max}]$ are delay and Doppler indices corresponding to the $i$-th path, where $l_{\max}$ and $k_{\max}$ denote the maximum delay and Doppler indices, respectively. $\kappa_i \in [-\frac{1}{2},\frac{1}{2}]$ is fractional Doppler shifts from nearest Doppler. The terms $\frac{1}{M \Delta f}$ and $\frac{1}{N T}$ represent the delay and Doppler resolutions, respectively. It is noteworthy that the sampling time $\frac{1}{M \Delta f}$ generally assumes a sufficiently small value, then the effects of fractional delays can be disregarded in typical wide-band systems \cite{liu2023predictive}.\par
At the receiver side, the received symbols in the DD domain can be obtained via
\begin{align}
    y_{\mathrm{DD}}[k,l]=\frac{1}{\sqrt{NM}}\sum_{n=0}^{N-1}\sum_{m=0}^{M-1}y_{\mathrm{TF}}[m,n]e^{-j2\pi (\frac{nk}{N}-\frac{ml}{M})},
    \label{receive_y_scalar}
\end{align}
where $y_{\mathrm{TF}}[m,n]$ denotes the received symbol in the discrete-TF domain.

\subsection{Input-Output Relationship Based on Bi-orthogonal Waveform}

In this section, we first consider that the transmit and receive pulse shaping waveforms satisfy the bi-orthogonal property. Therefore, the input-output relationship without the noise term in the DD domain can be written as \cite{wei2021waveform} 
\begin{align}
y_{\mathrm{DD}}[k,l] & =\frac{1}{NM}\sum_{k^{\prime}=0}^{N-1}\sum_{l^{\prime}=0}^{M-1}h_w[k^{\prime},l^{\prime}]x_{\mathrm{DD}}[k-k^{\prime},l-l^{\prime}].
\label{io_conv}
\end{align}
Here $h_w[\cdot, \cdot]$ denotes the sampled version of the impulse response function, which is given by 
\begin{align}
    h_w[k', l']=h_w(\nu, \tau)|_{\nu=\frac{k'}{NT}, \tau=\frac{l'}{M\Delta f}}.
\end{align}
Here, $h_w(\nu, \tau)$ is the circular convolution of the channel response with the SFFT of a rectangular windowing function in the TF domain, which is given by
\begin{align}
    h_{w}(\nu,\tau)=\int\int h(\tau^{\prime},\nu^{\prime})w(\nu-\nu^{\prime},\tau-\tau^{\prime})e^{-j2\pi\nu^{\prime}\tau^{\prime}}d\tau^{\prime}d\nu^{\prime},
    \label{sample_channel}
\end{align}
with 
\begin{align}
w(\nu,\tau)=\sum_{n=0}^{N-1}\sum_{m=0}^{M-1} e^{-j2\pi(\nu nT-\tau m\Delta f)}.
\end{align}
By substituting (\ref{ideal_channel}) into (\ref{sample_channel}), we obtain that 
\begin{align}
h_{w}(\nu,\tau) & =\sum_{i=1}^{P}h_{i}e^{-j2\pi\nu_{i}\tau_{i}}w(\nu-\nu_{i},\tau-\tau_{i}) \nonumber\\
 & =\sum_{i=1}^{P}h_{i}e^{-j2\pi\nu_{i}\tau_{i}}g_1(\tau,\tau_{i})g_2(\nu,\nu_{i}),
\end{align}
where $g_1(\tau,\tau_{i})\triangleq\sum_{m^{\prime}=0}^{M-1}e^{j2\pi(\tau-\tau_{i})m^{\prime}\Delta f}$, $g_2(\nu,\nu_{i})\triangleq\sum_{n^{\prime}=0}^{N-1}e^{-j2\pi(\nu-\nu_{i})n^{\prime}T}$. Let us analyze $g_1(\tau,\tau_{i})$ with sampled version as 
\begin{align}
   g_1(\tau,\tau_{i})|_{\tau=\frac{l^{\prime}}{M\Delta f}}=\sum_{m^{\prime}=0}^{M-1}e^{j\frac{2\pi}{M}(l^{\prime}-l_i)m^{\prime}} = \frac{e^{j2\pi(l^{\prime}-l_i)}-1}{e^{j\frac{2\pi}{M}(l^{\prime}-l_i)}-1}.
\end{align}
It can be observed that $g_1(\frac{l^{\prime}}{M\Delta f},\tau_{i}) = M$, for $[l'-l_i]_M=0$, and $g_1(\frac{l^{\prime}}{M\Delta f},\tau_{i}) = 0$ for the rest $[l'-l_i]_M$. Similarly, the sampled version of $g_2(\nu,\nu_{i})$ is given by
\begin{align}
    g_2(\frac{k'}{NT},\nu_{i})=\frac{e^{j2\pi(k^{\prime}-k_i- \kappa_i)}-1}{e^{j\frac{2\pi}{N}(k^{\prime}-k_i- \kappa_i)}-1}.
\end{align}
The magnitude of $\frac{1}{N}g_2(\frac{k'}{NT},\nu_{i})$ is given by 
\begin{align}
     \left|\frac{1}{N}g_2(\frac{k^{\prime}}{NT},\nu_{i}) \right| = \left|\frac{\sin (\pi(k'-k_i- \kappa_i))}{N \sin (\frac{\pi}{N}(k'-k_i- \kappa_i))}\right|.
     \label{fractional_Doppler}
\end{align}

As stated in \cite{wei2021transmitter}, the magnitude decreases rapidly with slope of ${\pi}(k'-k_i- \kappa_i)$. Here, we only consider $2N_i+1$ spread numbers around the peak $k_i$, i.e., $[k_i-N_i]_M \leq k' \leq [k_i+N_i]_M$, where $N_i$, $N_i \ll N$, denotes the truncated number indicating the parts that contain most of the energy. According to the approximation, the effective channel can be written as  
\begin{align}
     h_w[k', l'] \approx & \sum_{i=1}^{P} \sum_{q=-N_i}^{N_i}h_{i}e^{-j2\pi\nu_{i}\tau_{i}}g_1(\frac{l^{\prime}}{M\Delta f},\tau_{i}) \nonumber \\ &g_2(\frac{k^{\prime}}{NT},\nu_{i})\delta(k'-k_i-q).
     \label{sampled_hw}
\end{align}

\begin{figure}[tb]
    \centering
    \includegraphics[width=0.7\linewidth]{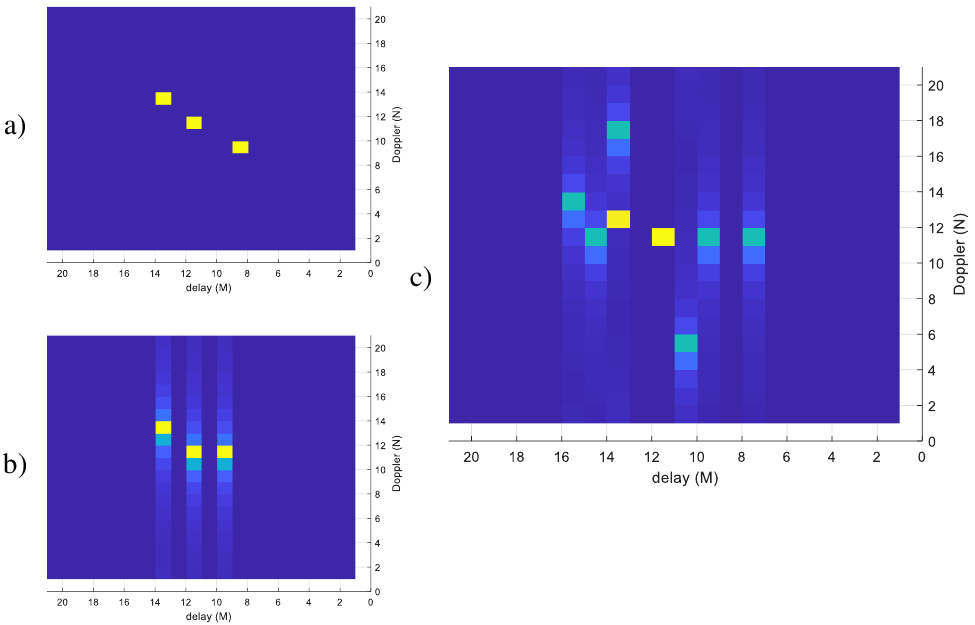}
    \caption{Different priors of OTFS channels due to different channel parameters, illustrating burst sparsity and block sparsity.}
    \label{OTFS_priors}
\end{figure}

Let us analyze the sparsity of $h_w[k',l']$ under different channel conditions. {In the scenario of integer delay and Doppler, i.e., $\kappa_i \approx 0$, which can be viewed as a reasonable approximation under low-mobility or grid-based bi-orthogonal pulses, the magnitude of $h_w[k',l']$ can be expressed as 
\begin{align}
    h_w[k',l'] = \sum_{i=1}^{P}h_{i}e^{-j2\pi\nu_{i}\tau_{i}}\delta(l'-l_i)\delta(k'-k_i).
\end{align}
}
The channel response is completely sparse in the DD domain, as depicted in Fig. \ref{OTFS_priors} (a), enabling efficient signal processing in this domain. In this case, the channel prior can be represented using the $\ell_1$-norm. {In the scenario of fractional Doppler, which typically arises in high mobility or with non-ideal pulses and leads to energy spreading across neighboring bins, the magnitude of $g_2(\nu,\nu_i)$ is given by (\ref{fractional_Doppler}).} As shown in Fig. \ref{OTFS_priors} (b), it can be observed that the sparsity decreases when the Doppler spread is large since signals are not compressed in the Doppler domain \cite{zhang2024wireless}. It is noteworthy that the degree of expansion is significantly influenced by the absolute value of $\kappa_i$. The sparsity of the channel can also be influenced by other variables, such as the number of paths $P$, as shown in Fig. \ref{OTFS_priors} (c). In the latter two cases, the channel exhibits block sparsity, making it challenging to effectively capture the channel priors using only the $\ell_1$ norm. Therefore, the input-output relationship in the DD domain can be written as \cite{Fei2021Message}
\begin{align}
  y_{\mathrm{DD}}[k,l]=\sum_{i=1}^{P}&\sum_{q=-N_i}^{N_i}h_ie^{-j2\pi\frac{\tau_i\nu_i}{MN}}\eta(q,\kappa_{i})\nonumber\\ & x_{\mathrm{DD}}[[k-k_i+q]_N, [l-l_i]_M]+v_{\mathrm{DD}}[k,l],
  \label{in_out_relationship}
\end{align}
where $\eta(q,\kappa_{i})=\frac{1-e^{-j2\pi(-q-\kappa_i)}}{N-Ne^{-j\frac{2\pi}{N}(-q-\kappa_i)}}$. The term $v_{\mathrm{DD}}[k,l]$ denotes the Gaussian noise in the DD domain.

\section{Problem Formulation}
\subsection{Problem Formulation for Channel Estimation}
\begin{figure}[tb]
    \centering
    \includegraphics[width=0.6\linewidth]{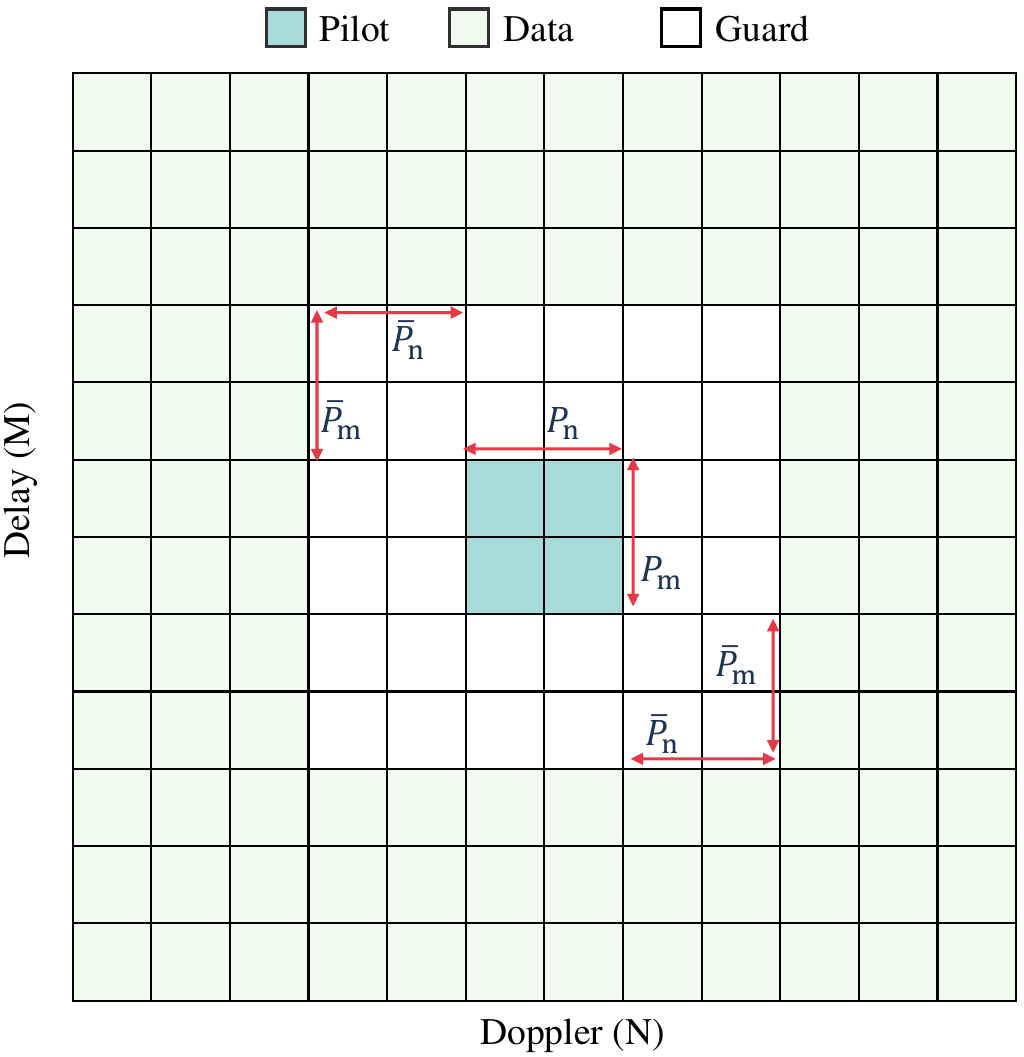}
    \caption{The considered OTFS pilot placement scheme, showing pilots in the DD domain surrounded by guarding interval with zeros.}
    \label{OTFS_P_r}
\end{figure}

{In this paper, the pilot scheme in \cite{wang2022joint} is adopted, where pilot symbols are placed in the DD domain, surrounded by guard space. As depicted in Fig. \ref{OTFS_P_r}, the lengths of pilots along the delay and Doppler are $P_{m}$ and $P_{n}$, respectively. {It is noteworthy that the deployment of multiple pilots can reduce peak-to-average power ratio (PAPR). In addition, the parameters $P_m$ and $P_n$ define a flexible pilot geometry, which can be configured to represent the conventional single-pilot pattern ($P_m=P_n=1$) or various multi-pilot structures depending on system requirements. Moreover, to mitigate interference between pilots and symbols, the length of guard intervals should be $\bar{P}_m \geq l_{\max}$ and $\bar{P}_n \geq 2\hat{k}_{\max}$, with $\hat{k}_{\max}$ being $k_{\max}+N_i$. Therefore, the pilot-plus-guard size equals $(P_m + 2l_{\max} )\times (P_n+4\hat{k}_{\max})$.}} \par 

{While grid-based bi-orthogonality on the sampled DD lattice is feasible with recent designs, in this work we adopt rectangular pulses, as they constitute a simple and widely used baseline that facilitates system implementation.} Based on this, the input-output relationship can be expressed as \cite{Fei2021Message}
\begin{align}
y_{\mathrm{DD}}[k,l] & =\sum_{i=1}^{P}\sum_{q=-N_{i}}^{N_{i}}h_{i}e^{j2\pi\left(\frac{l-l_{i}}{M}\right)\left(\frac{k_{i}+\kappa_{i}}{N}\right)}\alpha_{i}(k,l,q) \nonumber\\
 & \times x_{\mathrm{DD}}[[k-k_{i}+q]_{N},[l-l_{i}]_{M}]+\bar{n}[k,l],
 \label{input-output bi-waveform}
\end{align}
where 
\begin{align}
\alpha_i(k,l,q) =
\begin{cases} 
\eta(q,\kappa_{i}) & l_i \leq l < M, \\
(\eta(q,\kappa_{i})-\frac{1}{N})e^{-j2\pi\frac{[k-k_i+q]_N}{N}} & 0 \leq l < l_i,
\end{cases}
\end{align}
Based on the employed pilot placement scheme, we place the pilot symbols to ensure that their index satisfies $l \geq l_{\max}$, then (\ref{input-output bi-waveform}) can be rewritten as 
\begin{align}
y_{\mathrm{DD}}[k,l] & =\sum_{i=1}^{P}\sum_{q=-N_{i}}^{N_{i}}h_{i}\eta(q,\kappa_{i})e^{j2\pi\frac{(l-l_{i})(k_{i}+\kappa_{i})}{MN}} \nonumber \\
 & \times x_{\mathrm{DD}}\left[[k-k_{i}+q]_{N},[l-l_{i}]_{M}\right]+\bar{n}[k,l].
\label{input-output rect-waveform}
\end{align}\par 
For an arbitrary form of the DD domain signal model, e.g., (\ref{input-output rect-waveform}), here, we can rewrite it in a matrix form as
\begin{align}
\mathbf{y}_{\mathrm{DD}}=\boldsymbol{\Phi}_{\mathrm{DD}}\mathbf{h}+\bar{\mathbf{n}}_{\mathrm{DD}},
\label{prob_channel}
\end{align}
where $\mathbf{y}_{\mathrm{DD}} \in \mathbb{C}^{L_m\times 1},\bar{\mathbf{n}}_{\mathrm{DD}} \in \mathbb{C}^{L_m\times 1}$ with $L_m=(P_{m}+l_{\max}+1)(P_{n}+2\hat{k}_{\max}+1)$, $L_n=(l_{\max}+1) (2\hat{k}_{\max}+1)$. The measurement matrix $\boldsymbol{\Phi}_{\mathrm{DD}} \in \mathbb{C}^{L_m\times L_n}$ is derived from the pilot symbols with the $(k_i\, j_i)$-st row given by $\left [\mathbf{x}_{\mathrm{DD}}[k_i:k_i-2\hat{k}_{\max},\, j_i:j_i-l_{\max}]\right ]$, where $k_i \in \{0:P_n+2\hat{k}_{\max}\}$ and $j_i \in \{0:P_m+l_{\max}\}$. Since the additional phase term $e^{j2\pi\frac{l(k_{i}+\kappa_{i})}{MN}}$ in (\ref{input-output rect-waveform}) is related to both the channel and data, and $\mathbf{h}$ cannot be directly recovered when the measurement matrix $\boldsymbol{\Phi}_{\mathrm{DD}}$ contains unknown parameters, we neglect the additional phase difference with minimal performance degradation, as mentioned in \cite{shen2022random}. The term $\mathbf{h}\in \mathbb{C}^{L_n\times 1}$ can be inversely vectorized to obtain a truncated 2-dimensional (2D) DD channel $\mathcal{H}\in \mathbb{C}^{(l_{\max}+1)\times (2\hat{k}_{\max}+1)}$. It is noteworthy that the truncated DD channel $\mathcal{H}$ has finite support $[-\hat{k}_{\max}: \hat{k}_{\max}]$ along the Doppler dimension and $[0:l_{\max}]$ along the delay dimension, as specified in (\ref{sampled_hw}). For the vector $\mathbf{h}=[h_1,\cdots,h_{L_n}]$, the support set of $\mathbf{h}$ is defined as $\mathrm{supp}(\mathbf{h})=\{i:h_i\neq 0\}$, and $\mathbf{h}$ is \textit{$\gamma$-sparse} if $\left|\mathrm{supp}(\mathbf{h})\right|\leq \gamma$. \par 
As discussed earlier, the sparsity $\gamma$ varies due to the fractional spread and the increase in the number of paths. Traditionally, specific priors are required for algorithm design in different scenarios, such as the hidden Markov model (HMM) for message passing-based methods and $\ell_1$-norm priors for sparse Bayesian algorithms \cite{xiaoqi2024sparse, li2022uamp}. Leveraging the powerful capabilities of DL, we propose a unified PnP framework with a deep denoiser prior to address the channel estimation issue, thereby enhancing scalability.

\subsection{Problem Formulation for Symbol Detection}
To formulate the symbol detection problem, we derive the matrix form of the effective channels in the TF domain and the DD domain. Let $\mathbf{x}_{\mathrm{DD}}$ denote the 1D vector form of $\mathbf{X}_{\mathrm{DD}} \in \mathbb{Q}^{M \times N}$ obtained by column-wise rearrangement, i.e., $\mathbf{x}_{\mathrm{DD}} = \mathrm{vec}(\mathbf{X}_{\mathrm{DD}})$, where the $(k,l)$-th element $x_{\mathrm{DD}}[k,l]$ represents the DD information symbols. Similarly, $\mathbf{y}_{\mathrm{DD}}$ denotes the vector form of received symbols in the DD domain, $\mathbf{x}_{\mathrm{T}}$ is the transmitted vector in the time domain, and $\mathbf{y}_{\mathrm{T}}$ is the vector at the received side in the time domain. We utilize zero padding (ZP) \footnote{{The choice of padding does not alter the applicability of the proposed PnP framework. In particular, employing a cyclic prefix (CP) results in a circulant channel representation that allows lower-complexity implementation, while in this work, we adopt ZP as a general form for analysis.}} to implement the OTFS system with a padding length $L_{\mathrm{zp}} \geq l_{\max}$. After removing the ZP, the time domain input-output relation can be written as \cite{liu2023predictive}
\begin{align}
\mathbf{y}_\mathrm{T}=\mathbf{H}_{\mathrm{T}}\mathbf{x}_\mathrm{T}+\mathbf{n}_\mathrm{T},
\end{align}
where $\mathbf{n}_\mathrm{T}$ denotes the noise in the time domain. The effective channel in the time domain is a block diagonal matrix and given by $\mathrm{diag\{\mathbf{H}_1,\cdots, \mathbf{H}_n,\cdots, \mathbf{H}_N\}}$, where $\mathbf{H}_n\in \mathbb{C}^{M\times M}$ with $n\in [1,N]$ is the $n$-th sub-block effective channel matrix in the time domain. With the finite Fourier transform (FFT) and symplectic FFT operations, an equivalent vectorized form in the DD domain is given by \cite{liu2023predictive}
\begin{align}
\mathbf{y}_{\mathrm{DD}}=\mathbf{H}_{\mathrm{DD}}\mathbf{x}_{\mathrm{DD}}+\mathbf{n}_{\mathrm{DD}},
\label{symbol_dec}
\end{align}
where $\mathbf{n}_{\mathrm{DD}} \in \mathbb{C}^{MN\times 1}$ denotes the noise vector with each element following distribution $\mathcal{CN}(0, \sigma^2_\mathrm{d})$ and $\mathbf{H}_{\mathrm{DD}} \in \mathbb{C}^{MN\times MN}$ is the effective channel matrix in the DD domain, which is given by 
\begin{align}
\mathbf{H}_{\mathrm{DD}}=\left(\mathbf{F}_{N}\otimes\mathrm{G}_\mathrm{rx}\right)\mathbf{H}_{\mathrm{T}}\left(\mathbf{F}_{N}^{\mathrm{H}}\otimes\mathrm{G}_\mathrm{tx}\right).
\label{receive_y_vector}
\end{align}
In (\ref{receive_y_vector}), the matrices $\mathrm{G}_\mathrm{tx}=\mathrm{G}_\mathrm{rx}=\mathbf{I}_{\mathrm{M}}$ represent the matrix forms of the transmit and receive pulse shaping waveforms, respectively. To address the symbol detection problem, we combine an MLP for the PnP framework with a model-based time-domain equalization method to reduce computational complexity.

\section{Deep Learning-Based PnP Framework} 
This section introduces a novel PnP framework tailored for OTFS receiver design. Traditional OTFS receiver design often involves solving linear inverse problems, such as channel estimation (\ref{prob_channel}) and symbol detection (\ref{symbol_dec}), where the effectiveness of the solution typically depends on crafted priors that require manual tuning based on specific channel conditions. Our proposed PnP framework offers a powerful and flexible approach to tackle these challenges by alternately updating estimated parameters through a combination of model-based matrix computation methods and data-driven DL approaches. In the following, we elaborate on the architecture and application of this framework for OTFS receivers.

\begin{figure*}[tb]
    \centering
    \includegraphics[width=0.95\linewidth]{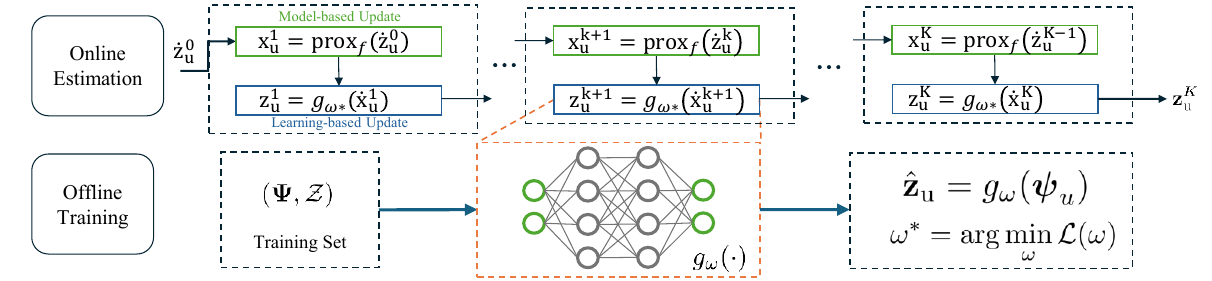}
    \caption{Illustration of the proposed PnP framework based on deep learning.}
    \label{train_p}
\end{figure*}

\subsection{The Unified Framework for OTFS Receivers}
To begin with, we consider a general linear inverse problem in signal processing, where the observed data $\mathbf{y}_\mathrm{u}$ can be modeled as
\begin{align}
\mathbf{y}_\mathrm{u}=\boldsymbol{\Phi}_\mathrm{u}\mathbf{x}_\mathrm{u}+\mathbf{n}_\mathrm{u},
\end{align}
where $\boldsymbol{\Phi}_\mathrm{u}\ \in \mathbb{C}^{a \times b}$ is the measurement matrix, and $\mathbf{n}_\mathrm{u} \in \mathbb{C}^{a \times 1}$ is a complex Gaussian noise vector. Here, $a$ and $b$ are arbitrary positive integers. {Prior knowledge of the unknown vector $\mathbf{x}_\mathrm{u}$, such as a detailed function representing the sparsity of $\mathbf{x}_\mathrm{u}$, is denoted by $\mathcal{J}(\mathbf{x}_\mathrm{u})$, where a conventional choice is the $\ell_1$-norm that corresponds to a Laplace or hierarchical Bayesian prior.} The objective is to recover $\mathbf{x}_\mathrm{u}$, which can be achieved by addressing the following problem, i.e., 
\begin{align}
    \min_{\mathbf{x}_{\mathrm{u}}} \frac{1}{2}||\mathbf{y}_\mathrm{u}-\boldsymbol{\Phi}_\mathrm{u}\mathbf{x}_\mathrm{u}||^2+\lambda \mathcal{J}(\mathbf{x}_{\mathrm{u}}),
    \label{origin problem}
\end{align}
where $\lambda \gg 0$ is a regularization parameter that balances the data-fidelity term and the regularization term. \par 
Proximal algorithms are widely used to solve optimization problems of the form presented in Eq. (\ref{origin problem}), especially when the regularization function $\mathcal{J}(\cdot)$ is non-smooth. {The deep PnP framework is rooted in proximal optimization methods, where alternating direction method of multipliers (ADMM) and related splitting algorithms efficiently decouple the data-fidelity and prior terms \cite{ryu2019plug}, thereby enabling flexibility in incorporating different priors.} The core idea of the deep PnP approach is to decouple the complex optimization problem into two simpler subproblems: one focusing on data fidelity and the other on incorporating prior knowledge. This decoupling is achieved by introducing an auxiliary variable $\mathbf{z}_\mathrm{u}$, transforming problem (\ref{origin problem}) into a constrained optimization problem $\mathbf{z}_\mathrm{u}$ as 
\begin{align}
    \min_{\mathbf{{x}_\mathrm{u}}}\quad &\frac{1}{2}\left\|\mathbf{\mathbf{y}_\mathrm{u}}-\boldsymbol{\Phi}_\mathrm{u}\mathbf{x}_\mathrm{u}\right\|_{2}^{2}+\lambda\mathcal{J}(\mathbf{z}_\mathrm{u})\nonumber \\ s.t.\quad &\mathbf{z}_\mathrm{u}=\mathbf{x}_\mathrm{u}.
    \label{auxiliary}
\end{align}
Then we rewrite (\ref{auxiliary}) by forming the augmented Lagrangian
function as 
\begin{align}
    \mathcal{L}_{\rho}\left(\mathbf{x}_\mathrm{u}, \mathbf{z}_\mathrm{u}, \mathbf{u}_\mathrm{u}\right)=&\frac{1}{2}\left\|\mathbf{y}_\mathrm{u}-\boldsymbol{\Phi}_\mathrm{u}\mathbf{x}_\mathrm{u}\right\|_{2}^{2}+\lambda\mathcal{J}\left(\mathbf{z}_\mathrm{u}\right)\nonumber \\&+\frac{\rho}{2}\left\|\mathbf{x}_\mathrm{u}-\mathbf{z}_\mathrm{u}\right\|_{2}^{2}+\mathbf{u}_\mathrm{u}^H\left(\mathbf{x}_\mathrm{u}-\mathbf{z}_\mathrm{u}\right),
    \label{lagrangian_f}
\end{align}
where $\rho \gg 0$ is a penalty parameter and $\mathbf{u}_\mathrm{u}$ is the dual variable of $\mathbf{z}_\mathrm{u}$. It is noteworthy that parameters $\rho$ and $\lambda$ play a crucial role in the entire alternating iterative optimization process, significantly affecting its convergence. To ensure that $\mathbf{x}_\mathrm{u}$ and $\mathbf{z}_\mathrm{u}$ converge to a fixed point, the value of $\rho$, which represents the weight of the constraint term, must progressively increase during the iterative process \cite{Wan2024Multitask}. Optimization problem (\ref{lagrangian_f}) can be solved by iteratively solving the following three
subproblems with variables $\mathbf{x}_\mathrm{u}$, $\mathbf{z}_\mathrm{u}$ and $\mathbf{u}_\mathrm{u}$ in turn, which is expressed as
\begin{align}
    &\mathbf{x}_\mathrm{u}^{k+1}=\arg\min_{\mathbf{x}_\mathrm{u}}f(\mathbf{x}_\mathrm{u})+\frac{\rho}{2}\left\|\mathbf{x}_\mathrm{u}-\mathbf{\dot{z}}_\mathrm{u}^{k}\right\|_{2}^{2}, \label{up_x}\\ &\mathbf{z}_\mathrm{u}^{k+1}=\arg\min_{\mathbf{z}_\mathrm{u}}\mathcal{J}\left(\mathbf{z}_\mathrm{u}\right)+\frac{1}{\sigma ^{2}}\left\|\mathbf{z}_\mathrm{u}-\mathbf{\dot{x}}_\mathrm{u}^{k+1}\right\|_{2}^{2}, \label{up_z}\\
    &\mathbf{u}_\mathrm{u}^{k+1}= \mathbf{u}_\mathrm{u}^{k}+ (\mathbf{x}_\mathrm{u}^{k+1}-\mathbf{z}_\mathrm{u}^{k+1}),
    \label{up_u}
\end{align}
where $f(\mathbf{x}_\mathrm{u})=\frac{1}{2}\left\|\mathbf{y}_\mathrm{u}-\boldsymbol{\Phi}_\mathrm{u}\mathbf{x}_\mathrm{u}\right\|_{2}^{2}$, $\mathbf{\dot{z}}_\mathrm{u}^{k}=\mathbf{z}_\mathrm{u}^k-\mathbf{u}_\mathrm{u}^k$, $\mathbf{\dot{x}}_\mathrm{u}^{k+1}=\mathbf{x}_\mathrm{u}^{k+1}+\mathbf{u}_\mathrm{u}^{k}$, and $\sigma^2=\lambda/2\rho$. As shown in Fig. \ref{train_p}, the developed deep PnP framework consists of two update methods: a model-based method is employed to solve (\ref{up_x}), while a DL-based method is utilized to solve (\ref{up_z}). \par 
\textbf{Update} $\mathbf{x}_\mathrm{u}^{k+1}$: Subproblem (\ref{up_x}) represents a standard LS problem, for which we can derive a closed-form solution, i.e., 
\begin{align}
    \mathbf{x}_\mathrm{u}^{k+1} &=(\boldsymbol{\Phi}_\mathrm{u}^H\boldsymbol{\Phi}_\mathrm{u}+\rho\, \mathbf{I})^{-1}(\boldsymbol{\Phi}_\mathrm{u}^H\mathbf{y}_\mathrm{u}+ \rho\, \mathbf{\dot{z}}_\mathrm{u}^{k}). \label{closed_form}
\end{align}
It is noteworthy that solution (\ref{closed_form}) can be simplified under specific properties of the measurement matrix $\boldsymbol{\Phi}_\mathrm{u}$.\par 

\textbf{Update} $\mathbf{z}_\mathrm{u}^{k+1}$: Subproblem (\ref{up_z}) is precisely the definition of a proximal mapping function \cite{bauschke2017correction} and can be expressed as $\mathbf{z}_\mathrm{u}^{k+1} = \mathrm{prox}_{\mathcal{J}}(\mathbf{\dot{x}}_\mathrm{u}^{k+1})$. Furthermore, this proximal function can be insightfully reformulated as a maximum a posteriori (MAP) estimation problem as
\begin{align}
     \mathbf{z}_\mathrm{u}^{k+1}& = \arg \max_{\mathbf{z}_\mathrm{u}} p(\mathbf{\dot{x}}_\mathrm{u}^{k+1}|\mathbf{z}_\mathrm{u}) \nonumber \\
     & = \arg \min_{\mathbf{z}_\mathrm{u}} \left( -\log p(\mathbf{z}_\mathrm{u}|\mathbf{\dot{x}}_\mathrm{u}^{k+1}) -\log p(\mathbf{z}_\mathrm{u})\right),
     \label{MAP_z}
\end{align}
where $p(\cdot)$ represents the probability density function (pdf). By comparing (\ref{MAP_z}) and (\ref{up_z}), we observe that the update step for $\mathbf{z}_\mathrm{u}^{k+1}$ can be interpreted as a Gaussian denoising problem. In this context, $\mathbf{\dot{x}}_\mathrm{u}^{k+1}$ is the noisy input, $\mathbf{z}_\mathrm{u}^{k+1}$ is the "denoised" signal, and $\sigma^2$ represents the noise variance. Nevertheless, directly solving this problem analytically is challenging due to the need for explicit knowledge of the prior $\mathcal{J}(\mathbf{z}_\mathrm{u})$, i.e., $p(\mathbf{z}_\mathrm{u})$, which is often complex and difficult to model. {Furthermore, the parameter $\rho$ increases during the iterative process, implying that the noise variance $\sigma^2$ changes continuously. This dynamic nature necessitates a denoiser that is both robust to varying noise levels and adaptable to complex priors, and thus replacing the explicit prior with a DL-based denoiser within the data-fidelity framework provides a feasible solution, as also shown in Bayesian-inspired deep learning studies \cite{xiaoqi2024sparse}.} \par 

\subsection{Offline Pre-Training for the Denoiser}
Fig. \ref{train_p} illustrates the two stages of the proposed approach: online estimation and offline training. We can acquire the following set for offline training as
\begin{align}
    (\boldsymbol{\Psi},\mathcal{Z})=\{(\boldsymbol{\psi}_\mathrm{u}^{(1)}, \mathbf{z}_\mathrm{u}^{(1)}), \cdots, (\boldsymbol{\psi}_\mathrm{u}^{(N_s)}, \mathbf{z}_\mathrm{u}^{(N_s)})\},
\end{align}
where $\boldsymbol{\psi}_\mathrm{u}^{(i)}$ and $\mathbf{z}_\mathrm{u}^{(i)}$ are the input and the ground truth of the $i$th, $i \in \{1,2,\cdots, N_s\}$, training sample, respectively. In particular, we generate the noisy input as $\boldsymbol{\psi}_\mathrm{u}^{(i)}=\mathbf{z}_\mathrm{u}^{(i)}+\mathbf{n}_\mathrm{u}^{(i)}$ with $\mathbf{n}_\mathrm{u}^{(i)}\sim \mathcal{CN}(0,\sigma_n^2)$ being the additive noise. Note that the noise variance is uniformly distributed across a defined range, enabling the denoiser to handle varying noise levels effectively. Let $g_{\omega}(\cdot)$ represent the mapping function of a specific deep neural network (DNN) with $\omega$ being network parameters, and the network is trained by minimizing the following mean squared error (MSE) loss function 
\begin{align}
    \mathcal{L}(\omega)=\sum_i^{N_s} \frac{\|g_{\omega}(\boldsymbol{\psi}_\mathrm{u}^{(i)})-\mathbf{z}_\mathrm{u}^{(i)}\|^2_{2}}{\|\mathbf{z}_\mathrm{u}^{(i)}\|^2_{2}}.
\end{align}
Based on this, the DNN can utilize the backpropagation (BP) algorithm to iteratively update its network parameters, thereby optimizing the model. The optimal parameters $\omega^*$ are obtained by minimizing the loss function i.e., $\omega^* =\arg \min \mathcal{L}(\omega)$.\par
\vspace{-0.1cm}
\subsection{Online Estimation and Generalization}
For the online estimation phase, we have a different testing set as 
\begin{align}
    (\mathcal{\dot{Y}},\mathcal{\dot{X}})=\{(\mathbf{\dot{y}}_\mathrm{u}^{(1)}, \mathbf{\dot{x}}_\mathrm{u}^{(1)}), \cdots, (\mathbf{\dot{y}}_\mathrm{u}^{(N_s')}, \mathbf{\dot{x}}_\mathrm{u}^{(N_s')})\},
\end{align}
where $\mathbf{\dot{y}}_\mathrm{u}^{(i)}$ and $\mathbf{\dot{x}}_\mathrm{u}^{(i)}$ denote the input and ground truth of the $i$th testing sample, respectively, with $i \in \{1,2,\cdots, N_s'\}$. Here, $N_s'$ is the sample size. {Given the input $\mathbf{\dot{y}}_\mathrm{u}^{(i)}$, the final estimated $\hat{\mathbf{\dot{x}}}_\mathrm{u}^{(i)}$ is obtained by alternately updating the model-based step (\ref{closed_form}) and the data-driven step with the pre-trained denoiser. The forward operator is computed for each frame so that frame-to-frame variations in channel conditions are accounted for in the observation model. A significant advantage of the PnP framework is its flexibility in integrating different denoiser architectures, such as recurrent neural networks (RNNs), fully connected networks (FCNs), or convolutional neural networks (CNNs) \cite{lecun2015deep}. This flexibility allows a balance between performance and complexity, and in practice, the pre-trained denoiser can be directly applied during the online phase, with retraining required only if the underlying channel prior changes substantially.}

\section{OTFS Receiver Processing Using the PnP Framework}
In this section, we detail the application of our proposed PnP framework for the OTFS channel estimation and symbol detection, leveraging two specifically tailored deep neural networks. {The direct application of PnP frameworks to communication systems presents distinct challenges, as communication signals possess structural characteristics that necessitate tailored denoising strategies, while wireless systems simultaneously require high estimation accuracy and low-latency constraints.} To address these challenges, we introduce an EDN-based approach for channel estimation, designed to exploit the inherent sparsity and structural properties of the OTFS channel. At the same time, to meet the low-latency requirements of symbol detection, we simplify the inverse matrix computations and develop a dedicated network-based algorithm that incorporates constellation priors.

\subsection{PnP for Channel Estimation}
This subsection describes the proposed PnP-CE algorithm for the OTFS channel estimation task. Algorithm \ref{Al_CE} describes the EDN-based realization of the PnP framework, where $\mathbf{u}_\mathrm{h}$ is the dual variable of $\mathbf{z}_\mathrm{h}$. According to (\ref{prob_channel}), the OTFS channel estimation problem can be reformulated as 
\begin{align}
    \min_{\mathbf{h}} \frac{1}{2}||\mathbf{y}_\mathrm{DD}-\boldsymbol{\Phi}_\mathrm{DD}\mathbf{h}||^2+\lambda_\mathrm{h} \mathcal{J}_\mathrm{h}(\mathbf{h}),
    \label{channel problem}
\end{align}
where $\lambda_\mathrm{h}$ and $\mathcal{J}_\mathrm{h}(\cdot)$ denote the penalty parameter and channel prior, respectively. According to the PnP framework, the channel $\mathbf{h}$ can be estimated via (\ref{origin problem}). For the model-based update step, the $k+1$-th iteration of $\mathbf{h}$ is given by the closed-form solution: 
 \begin{align}
    \mathbf{h}^{k+1}&= \arg \min_{\mathbf{h}} \frac{1}{2}\left\|\mathbf{y}_\mathrm{DD}-\boldsymbol{\Phi}_\mathrm{DD}\mathbf{h}\right\|_{2}^{2}+\frac{\rho_\mathrm{h}}{2}\left\|\mathbf{h}-\mathbf{\dot{z}}_\mathrm{h}^{k}\right\|_{2}^{2}\nonumber \\
    &= (\boldsymbol{\Phi}_\mathrm{DD}^H\boldsymbol{\Phi}_\mathrm{DD}+\rho_\mathrm{h}\, \mathbf{I})^{-1}(\boldsymbol{\Phi}_\mathrm{DD}^H\mathbf{y}_\mathrm{DD}+ \rho_\mathrm{h}\, \mathbf{\dot{z}}_\mathrm{h}^{k}),
\end{align}
where $\mathbf{\dot{z}}_\mathrm{h}^{k}$ denotes the auxiliary variable, $\rho_\mathrm{h}$ is the penalty parameter for channel estimation. Note that PnP-CE can be used for arbitrary pilot patterns since the DL-based denoiser works for auxiliary variables and is decoupled from the configuration of the pilot patterns. The $(k+1)$-th iteration of $\mathbf{\dot{z}}_\mathrm{h}$ can be expressed as 
\begin{align}
\mathbf{z}_\mathrm{h}^{k+1} =  \arg\min_{\mathbf{z}_\mathrm{h}}\mathcal{J}_\mathrm{h}\left(\mathbf{z}_\mathrm{h}\right)+\frac{1}{\sigma_\mathrm{h} ^{2}}\left\|\mathbf{z}_\mathrm{h}-\mathbf{\dot{h}}^{k+1}\right\|_{2}^{2}.
\end{align}
Here, $\mathbf{\dot{h}}^{k+1} = \mathbf{h}^{k+1} + \mathbf{u}_\mathrm{h}^{k}$, where $\mathbf{u}_\mathrm{h}^{k}$ is the dual variable, and $\sigma_\mathrm{h} = \frac{\lambda_{\mathrm{h}}}{2\rho_\mathrm{h}}$. The channel prior $\mathcal{J}_\mathrm{h}$ varies across different scenarios, requiring distinct parameterizations, which presents a significant challenge for algorithm design. For example, in some cases, the prior can be modeled as a binary Gaussian distribution with integer delay and Doppler, while in other cases, the channel prior may follow a Markov chain model with fractional delay and Doppler \cite{li2022uamp}. To solve the channel estimation problem in various channel configurations and applications, we develop a DL-based denoiser to update the auxiliary variable $\mathbf{z}_\mathrm{h}^{k+1}$ as
\begin{align}
    \mathbf{z}_\mathrm{h}^{k+1} = \mathrm{prox}_{\mathcal{J}_\mathrm{h}}(\mathbf{\dot{h}}^{k+1}) = D_\mathrm{h}(\mathbf{\dot{h}}^{k+1}),
\end{align}
where $D_\mathrm{h}$ is the DL-based denoiser. To implement the denoiser, we adopt a lightweight EDN architecture due to its suitability for the characteristics of the DD domain channel. These channels typically exhibit sparsity, with a few dominant taps that are localized, while practical imperfections such as fractional delays and Doppler introduce energy spreading around these tap locations. The encoder-decoder structure of EDN effectively captures multi-scale features, identifying both the global positions of sparse components and their local spreading patterns.  \par 

\begin{figure}[tb]
    \centering
    \includegraphics[width=0.95\linewidth]{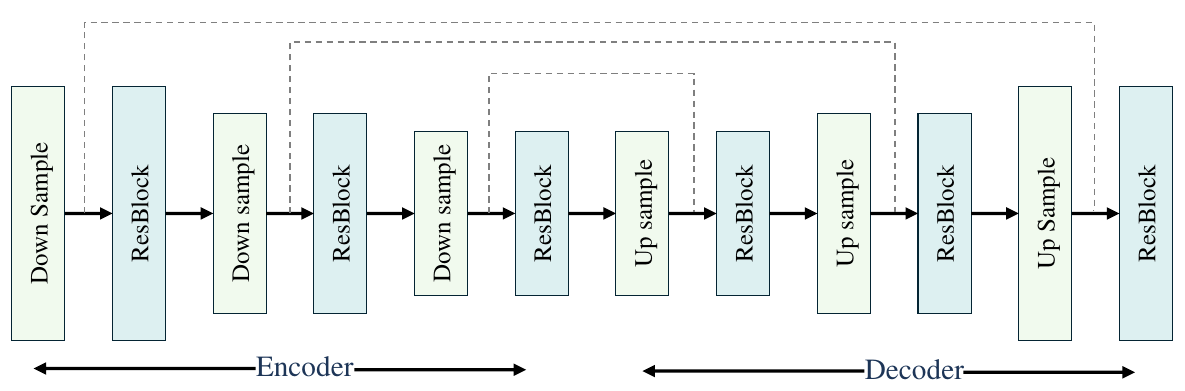}
    \caption{Block diagrams showing the deep encoder-decoder network.}
    \label{Unet}
\end{figure}

As depicted in Fig. \ref{Unet}, the lightweight EDN is composed of one input layer, one encoder block, one decoder block, and one output layer. The hyperparameters and the details of each layer of EDN are introduced as follows.
\subsubsection{Input and Output Layers} To extract features from the complex-valued input data $\boldsymbol{\psi}_\mathrm{h}$, we separate the input as the real part and the imaginary part as $\tilde{\boldsymbol{\psi}}_\mathrm{h} = \mathcal{M}(\mathrm{Re}\{\boldsymbol{\psi}_\mathrm{h}\},\mathrm{Im}\{\boldsymbol{\psi}_\mathrm{h}\}),$
where $\mathcal{M}(\cdot)$ represents a mapping function. Similarly, the output of the network is given by $\hat{\mathbf{z}}_\mathrm{h} = \mathcal{M}^{-1}(D_\mathrm{h}(\tilde{\boldsymbol{\psi}}_\mathrm{h}))$, 
where $\mathcal{M}^{-1}$ denotes a mapping function that constructs the complex-valued matrix from the real-valued components.

\subsubsection{Encoder and Decoder Layers} The EDN is composed of a series of downsample and upsample convolution blocks, which systematically reduce the spatial dimensions while increasing the number of feature channels. Each downsample convolution block incorporates $N_c$ sequential layers of $3 \times 3 \times 3$ convolution operations, which are subsequently processed by rectified linear unit (ReLU) activation and batch normalization (BN). As the network depth grows, the training process becomes increasingly susceptible to instability, making it challenging to achieve convergence to an optimal local minimum. To address this issue, we adopt a residual learning framework inspired by ResNet \cite{he2016deep}. Specifically, the basic block of our network is designed as a residual block, which integrates $3 \times 3 \times 3$ convolution units alongside a projection shortcut implemented through a $1 \times 1 \times 1$ convolution. This projection shortcut preserves the flow of shallow-level information, ensuring gradient propagation and mitigating the vanishing gradient problem. By combining these shortcuts with skip connections, our architecture facilitates both local and global information interactions, significantly enhancing the network's representational capacity. \par 
\begin{algorithm}[t]
\caption{PnP for OTFS channel estimation}
\centering
\resizebox{0.85\linewidth}{!}{%
\begin{minipage}{\linewidth}
\begin{algorithmic}[1]
\Require Transmitted pilot $\boldsymbol{\Phi}_\mathrm{DD}$, received signal $\mathbf{y}_\mathrm{DD}$, number of iterations $N_K$
\Ensure Estimated $\mathbf{h}^{N_K}$
\State Initialize auxiliary ${\mathbf{z}}_\mathrm{h}^0$, $\mathbf{u}_\mathrm{h}^{0}$ and ${\mathbf{h}}^0$
\For{$k = 1, 2, \dots, N_K$}
\State $\mathbf{h}^k=\mathrm{prox}_{f_\mathrm{h}}(\dot{\mathbf{z}}_\mathrm{h}^{k-1})$
\State $\mathbf{z}_{\mathrm{h}}^k=D_{\mathrm{h}}(\dot{\mathbf{h}}^k)$
\State $\mathbf{u}_\mathrm{h}^{k}= \mathbf{u}_\mathrm{h }^{k-1}+ (\mathbf{h}^{k}-\mathbf{z}_\mathrm{h}^{k})$
\EndFor 
\end{algorithmic}
\end{minipage}}
\label{Al_CE}
\end{algorithm}

\subsection{PnP for Symbol Detection}

This subsection describes the proposed PnP-SD algorithm for the symbol detection problem. The overall symbol detection method is detailed in Algorithm \ref{Al_SD}. Let us first model the channel estimation error as
\begin{align}
     \mathbf{H}_\mathrm{DD} = \hat{\mathbf{H}}_\mathrm{DD}+\mathbf{E}_\mathrm{DD},
\end{align}
where $\hat{\mathbf{H}}_\mathrm{DD}$ and $\mathbf{E}_\mathrm{DD}$ are the estimated channel matrix and estimation error, respectively. {Similar to \cite{yoo2004capacity}, $\hat{\mathbf{H}}_\mathrm{DD}$ and $\mathbf{E}_\mathrm{DD}$ are assumed independent, with $\mathbf{E}_\mathrm{DD}$ having zero-mean CSCG entries of variance $\frac{1}{MN}\sigma_e^2$. Therefore, the symbol detection model can be written as 
\begin{align}
\mathbf{y}_\mathrm{DD}&=\hat{\mathbf{H}}_\mathrm{DD}\mathbf{x}_{\mathrm{DD}}+\mathbf{E}_\mathrm{DD}\mathbf{x}_{\mathrm{DD}}+\mathbf{n}_{\mathrm{DD}}\nonumber \\&
=\hat{\mathbf{H}}_\mathrm{DD}\mathbf{x}_{\mathrm{DD}}+\hat{\mathbf{n}}_{\mathrm{DD}},
\end{align}
where the entries of $\hat{\mathbf{n}}_{\mathrm{DD}}$ are CSCG noise with variance $\sigma^2_\mathrm{d}+\sigma^2_e$ when $MN$ is large enough. To quantify imperfection, we define the normalized error $\epsilon=\frac{\sigma_e^2}{||\mathbf{H}_\mathrm{DD}||^2_F}$.} Accordingly, the OTFS symbol detection can be reformulated as 
\begin{align}
    \min_{\mathbf{x}_\mathrm{DD}} \frac{1}{2}||\mathbf{y}_\mathrm{DD}-\hat{\mathbf{H}}_\mathrm{DD}\mathbf{x}_\mathrm{DD}||^2+\lambda_\mathrm{x} \mathcal{J}_\mathrm{x}(\mathbf{x}_\mathrm{DD}),
    \label{symbol problem}
\end{align}
where $\lambda_\mathrm{x}$ and $\mathcal{J}_\mathrm{x}(\cdot)=\mathbf{x}_\mathrm{DD}\in \mathbb{Q}^{MN\times 1}$ denote the penalty parameter and symbol prior, respectively. \par 
Following the PnP framework, the model-based step for updating $\mathbf{x}_\mathrm{DD}^k$ is given by
\begin{align}
    \mathbf{x}_{\mathrm{DD}}^{k+1}&= \arg \min_{\mathbf{x}_\mathrm{DD}} \frac{1}{2}\left\|\mathbf{y}_\mathrm{DD}-\hat{\mathbf{H}}_\mathrm{DD}\mathbf{x}_\mathrm{DD}\right\|_{2}^{2}+\frac{\rho_\mathrm{x}}{2}\left\|\mathbf{x}_\mathrm{DD}-\mathbf{\dot{z}}_\mathrm{x}^{k}\right\|_{2}^{2} \nonumber \\
    &= \mathbf{W}^{-1}(\hat{\mathbf{H}}_\mathrm{DD}^H\mathbf{y}_\mathrm{DD}+ \rho_\mathrm{x}\, \mathbf{\dot{z}}_\mathrm{x}^{k}),
    \label{decouple channel}
\end{align}
where $\mathbf{W}=(\hat{\mathbf{H}}_\mathrm{DD}^H\hat{\mathbf{H}}_\mathrm{DD}+\rho_\mathrm{x}\, \mathbf{I}_{\mathrm{MN}})$ and $\rho_\mathrm{x}$ denotes the penalty parameter. To reduce the computational complexity of symbol detection, we perform $\mathbf{W}^{-1}$ in the time domain. According to (\ref{receive_y_vector}), the inverse matrix can be written as 
\begin{align}
\mathbf{W}^{-1}&=\left(\mathbf{F}_{N}\otimes\mathbf{I}_\mathrm{M}\right)\left(\hat{\mathbf{H}}^{H}_{\mathrm{T}}\hat{\mathbf{H}}_{\mathrm{T}}+\rho_\mathrm{x}\mathbf{I}_{\mathrm{MN}}\right)^{-1}\left(\mathbf{F}_{N}^{\mathrm{H}}\otimes\mathbf{I}_\mathrm{M}\right),
\end{align}
where $\hat{\mathbf{H}}_{\mathrm{T}}$ denotes the estimated channel matrix in the time domain. Thanks to the block-diagonal structure of $\hat{\mathbf{H}}_{\mathrm{T}}$, we can perform the matrix inversion with low computational complexity by inverting the submatrices, i.e., \cite{haifeng2022cross}
\begin{align}
\left(\hat{\mathbf{H}}^{H}_{\mathrm{T}}\hat{\mathbf{H}}_{\mathrm{T}}+\rho_\mathrm{x}\mathbf{I}_{\mathrm{MN}}\right)^{-1}=\operatorname { diag}\{\mathbf{C}^{-1}_{1}, \cdots, \mathbf{C}^{-1}_{N}\},
\end{align}
where $\mathbf{C}_n=\left({\hat{\mathbf{H}}_n^H\hat{\mathbf{H}}_n+\rho_\mathrm{x} \mathbf{I}_{\mathrm{M}}}\right) \in \mathbb{C}^{M \times M }$  are the diagonal sub-block of channel matrix $\mathbf{W}$. \par 

\begin{algorithm}[tb]
\caption{PnP for OTFS symbol detection}
\centering
\resizebox{0.85\linewidth}{!}{%
\begin{minipage}{\linewidth}
\begin{algorithmic}[1]
\Require Estimated channel $\hat{\mathbf{H}}_\mathrm{DD}$, received signal $\mathbf{y}_\mathrm{DD}$, number of iterations $N_K$ 
\Ensure Estimated symbol $\mathbf{x}_\mathrm{DD}^{N_K}$
\State Initialize auxiliary $\dot{\mathbf{z}}_\mathrm{x}^0$, $\mathbf{u}_\mathrm{x}^{0}$ and ${\mathbf{x}}_{\mathrm{DD}}^0$
\For{$k = 1, 2, \dots, N_K$}
\State $\mathbf{x}_\mathrm{DD}^k=\mathrm{prox}_{f_\mathrm{x}}(\dot{\mathbf{z}}_\mathrm{x}^{k-1})$
\State $\mathbf{z}_\mathrm{x}^k=D_{\mathrm{x}}(\dot{\mathbf{x}}_\mathrm{DD}^k)$
\State $\mathbf{u}_\mathrm{x}^{k}= \mathbf{u}_\mathrm{x}^{k-1}+ (\mathbf{x}_\mathrm{DD}^{k}-\mathbf{z}_\mathrm{x}^{k})$
\EndFor 
\end{algorithmic}
\end{minipage}}
\label{Al_SD}
\end{algorithm}

\begin{figure}[tb]
    \centering
    \includegraphics[width=0.95\linewidth]{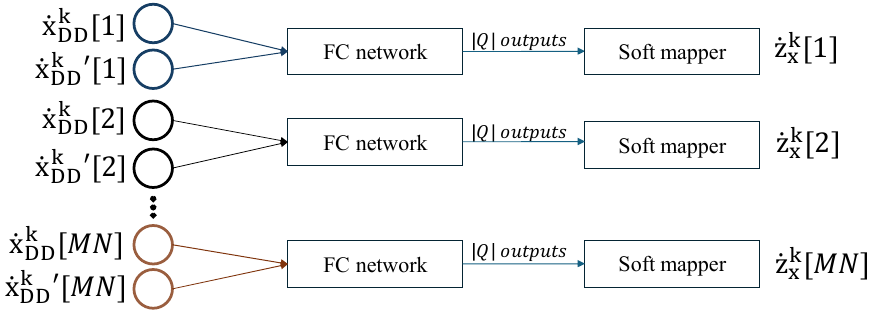}
    \caption{The proposed MLP-denoiser, with 3 hidden layers and one softmax activation layer.}
    \label{Symbol_LSTM}
\end{figure}

Since the decoupling operation is already performed in (\ref{decouple channel}), the update for auxiliary variable $\mathbf{\dot{z}}_\mathrm{x}^{k}$ becomes a denoising task. For the learning-based denoiser, we employ an MLP-based detection method and soft information technology to obtain the auxiliary variable. Based on the type of constellation, we propose updating the auxiliary variables $\mathbf{z}_\mathrm{x}$ as 
\begin{align}
    \mathbf{z}_\mathrm{x}^{k+1} =\arg\min_{\mathbf{z}_\mathrm{x}}\mathcal{J}_\mathrm{x}\left(\mathbf{z}_\mathrm{x}\right)+\frac{1}{\sigma_\mathrm{x} ^{2}}\left\|\mathbf{z}_\mathrm{x}-\mathbf{\dot{x}}_\mathrm{DD}^{k+1}\right\|_{2}^{2}= D_\mathrm{x}(\mathbf{\dot{x}}_\mathrm{DD}^{k+1}),
    \label{up_symbol_x}
\end{align}
where $\sigma_\mathrm{x}=\frac{\lambda_{\mathrm{x}}}{2\rho_\mathrm{x}}$, and $D_\mathrm{x}(\cdot)$ represents the proposed MLP update denoiser, which is depicted in Fig. \ref{Symbol_LSTM}. The $2MN$ input neurons are fed to the MLP, which consists of 3 hidden layers and one softmax activation layer. The input neurons, consisting of the real and imaginary parts of the input vector $\mathbf{\dot{x}}_\mathrm{DD}^{k+1}$, are processed by the MLP and then passed through the final softmax layer to produce $Q'=|\mathbb{Q}|$ outputs. The $Q'$ output neurons correspond to the size of the modulation alphabet, with each neuron representing the probability of a particular constellation point in the alphabet. To further improve the detection accuracy, we exploit the constellation prior and update $\mathbf{z}_\mathrm{x}^{k+1}$ via a soft mapper as 
\begin{align}
   z^{k+1}[m'] = \sum_{q\in \mathbb{Q}}p(z_{\mathrm{x}}^{k+1}[m']=q | \dot{x}_{\mathrm{DD}}^{k+1}[m'])\times q,
\end{align}
where $\dot{x}_{\mathrm{DD}}^{k+1}[m']$ with $m' \in [1, MN]$ is the element of $\mathbf{\dot{x}}_\mathrm{DD}^{k+1}$. After completing MLP-based denoiser for all $MN$ symbols, we use (\ref{decouple channel}) to obtain $\mathbf{\dot{x}}_\mathrm{DD}^{k+2}$ at the $k+2$ iteration. In contrast to the completely data-based DL, we effectively utilize the estimated channel information and DL to achieve more accurate symbol detection results. \par
{As shown in Table \ref{tab:complex_time}, the complexity of the PnP-SD detector in one iteration is expressed as $\mathcal{O}\left(M^3N+I_L\sum_l^Lz_{l-1}z_{l})\right)$, where $I_L$ is the layer numbers of the MLP, $z_{l}$ is the neuron number at the $l$-th layer. In the offline training phase, we train an adaptive denoiser for the PnP framework with the complexity being $\mathcal{O}\left(I_tI_L\sum_l^Lz_{l-1}z_{l})\right)$, with $I_t$ indicating the number of training epochs. Compared to the conventional LS and LMMSE detectors, whose complexity is given by $\mathcal{O}\left(M^3N^3\right)$, the proposed detector exhibits significantly lower computational complexity. While the PnP-SD detector exhibits a relatively higher computational complexity compared to OAMP, its actual execution time is considerably shorter. This advantage arises because the neural network components within PnP-SD can be efficiently parallelized on modern GPUs.}
\begin{table}[tb]
    \centering
    \caption{{Computational Complexities and Running Times of Different Algorithms}}
    \resizebox{1\linewidth}{!}{
    \begin{tabular}{ccc} 
        \toprule
        \textbf{Algorithm} & \textbf{Offline Training (Complexity / Time)} & \textbf{Online Estimation (Complexity / Time)} \\
        \midrule
        LS & - & $\mathcal{O}(M^3N^3)$ / $0.0070\,\mathrm{s}$ \\
        \midrule
        LMMSE & - & $\mathcal{O}(M^3N^3)$ / $0.0079\,\mathrm{s}$ \\
        \midrule
        OAMP & - & $\mathcal{O}(MN+2MN\log_{2}(MN)+MN|\mathbb{Q}|)$ / $0.2082\,\mathrm{s}$ \\
        \midrule
        PnP-SD & $\mathcal{O}\!\left(I_tI_L\sum_{l=1}^L z_{l-1}z_{l}\right)$ / $518.2742\,\mathrm{s}$ & $\mathcal{O}\!\left(M^3N+I_L\sum_{l=1}^L z_{l-1}z_{l}\right)$ / $0.0028\,\mathrm{s}$ \\
        \bottomrule
    \end{tabular}}
    \label{tab:complex_time}
\end{table}

\section{Simulation Results}
\subsection{Parameters Setting}
An OTFS frame is configured with $N=20$ and $M=20$\footnote{{A reduced frame size is employed to ensure tractable implementation of all baselines while still preserving the essential DD structures. Although the resulting overhead appears relatively large in this setting, it diminishes rapidly with larger frames in practical scenarios (e.g., about $0.88\%$ when $M=1024$).}},  corresponding to 20 time slots and 20 subcarriers in the TF domain. {For channel estimation, we embed a $(P_m \times P_n)=(2 \times 2)$ multi-pilot block within the frame, surrounded by guard symbols.} {The system operates at a carrier frequency of $f_c=3$ GHz with a subcarrier spacing of 7.5 kHz. The maximum delay index and Doppler index are set as $k_{\max}=3$ and $l_{\max}=4$, respectively. The $k=\{1,2,3\}$ correspond approximately to vehicle speeds $\{135,\, 270,\, 405\} \;\mathrm{km/h}$, which is representative of high-speed rail scenarios.} The number of channel paths is $P=5$, and the associated delay and Doppler indices for each channel path are randomly selected from $[0, l_{\max}]$ and $[-k_{\max}, k_{\max}]$, respectively. The channel gain follows the distribution $\mathcal{CN}(0, \frac{1}{P})$. The signal-to-noise ratio (SNR) for the transmission is defined as $\mathrm{SNR}=\frac{E_s}{N_0}$, where 4QAM symbol energy $E_s$ is normalized to 1. Moreover, we generate the dataset using the Monte-Carlo method. The training set and validation set consist of $90,000$ and $10,000$ samples, respectively, which are employed to train the DL-based denoisers. The added random Gaussian noise for training the DL-based denoisers is uniformly generated in the range of $[0, 50]$ dB.  {For online estimation, each presented result is obtained by averaging over $2,000$ Monte-Carlo realizations, and all methods are evaluated under identical settings to ensure fair comparison.} Note that the Adam optimizer is adopted for network training, with the learning rate set to 0.001. For the PnP-based channel estimation and symbol detection methods, the regularization parameters $\lambda_\mathrm{x}=\lambda_\mathrm{h}$ and the penalty factors $\rho_\mathrm{x}=\rho_\mathrm{h}$ are tuned and set to $0.5$ and $0.1$, respectively.\par

\begin{figure}[tt]
    \centering
    \includegraphics[width=0.7\linewidth]{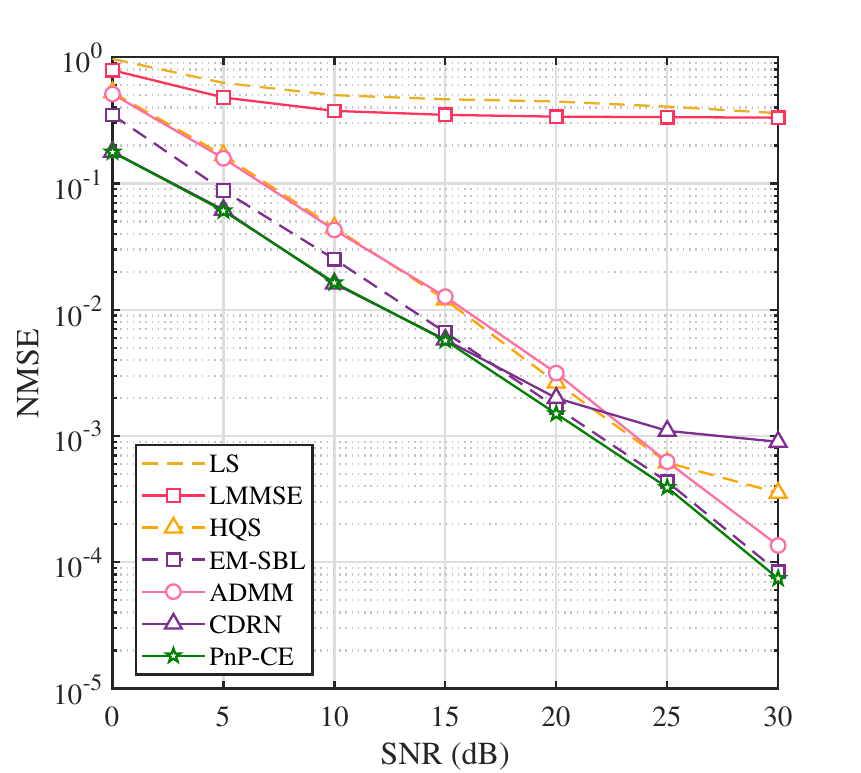}
    \caption{{The performance comparison for channel estimation.}}
    \label{Baseline_Channel}
\end{figure}

\begin{figure}[tb]
    \centering
    \includegraphics[width=0.65\linewidth]{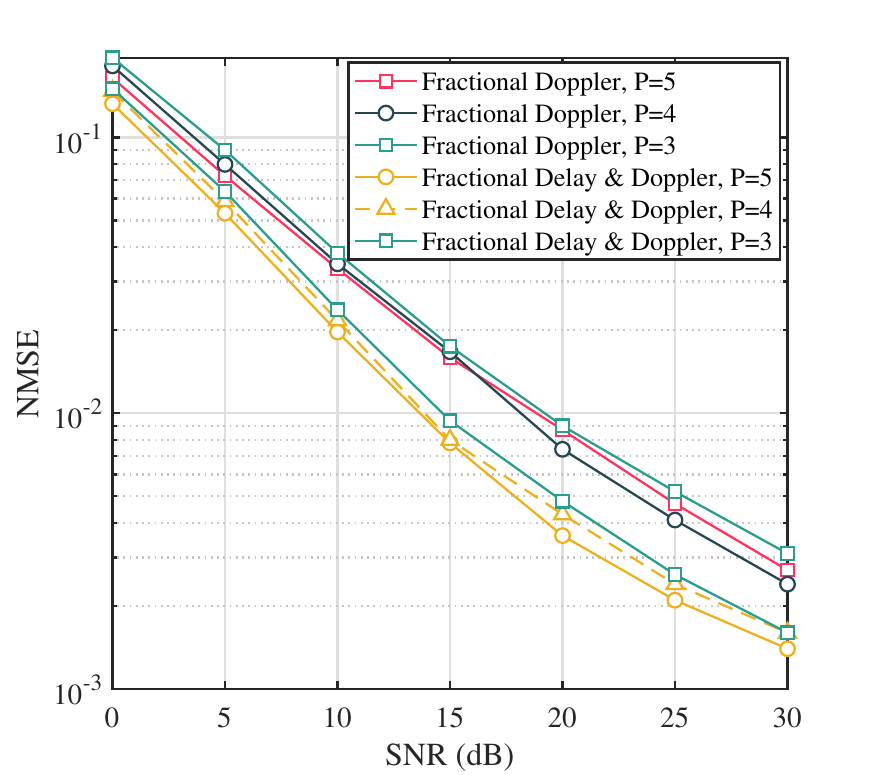}
    \caption{{NMSE versus SNR curves for channel estimation, illustrating the impact of fractional Doppler and fractional DD effects under different path numbers.} }
    \label{Fractional_Channel}
\end{figure}

\begin{figure}[tb]
    \centering
    \includegraphics[width=0.65\linewidth]{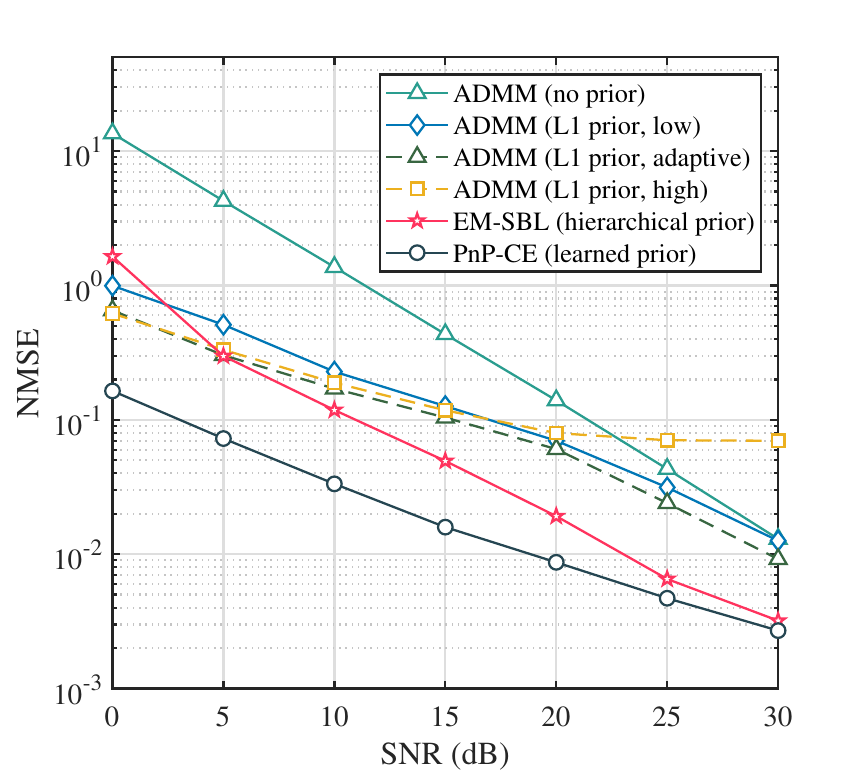}
    \caption{{NMSE versus SNR for channel estimation under different prior settings.}}
    \label{different_priors}
\end{figure}

\subsection{Results for Channel Estimation}
{We provide simulations of the following algorithms for comparison: the LS method, the LMMSE method, the half-quadratic splitting (HQS) method \cite{Wan2024Multitask}, the expected maximum (EM)-SBL \cite{Zhao2020SBL}, and the ADMM method \cite{yuan2018iterative}, and the convolutional deep residual network (CDRN) \cite{Liu2022RIS}.}  {The channel estimation performance under the integer delay and Doppler case, evaluated in terms of normalized mean square error (NMSE) across various SNR levels, is illustrated in Fig. \ref{Baseline_Channel}.} {A noticeable performance gap is observed between the LS and LMMSE methods. LS performs the worst since it ignores the channel sparsity and amplifies noise through the measurement operator, while LMMSE partly improves performance but suffers from bias and residual interference, leading to an NMSE floor at high SNR. HQS outperforms LS and LMMSE but degrades at high SNR due to approximation bias in quadratic splitting, whereas ADMM enforces the constraint more tightly and achieves better performance at 30 dB. The EM-SBL method performs well at medium-to-high SNR, where the hierarchical prior and EM-based updates effectively capture fractional spreading. However, the performance degrades at low SNR, where hyperparameter estimation becomes unreliable and error propagation leads to significant loss of accuracy. In comparison, the CDRN method demonstrates effective denoising capability at low SNR, while its performance saturates at higher SNR since it does not fully exploit the structured sparsity of the channel, and its limited generalization further restricts scalability. By contrast, the proposed PnP-CE consistently outperforms all baselines across the entire SNR range, maintaining robustness under both low- and high-SNR conditions. This advantage arises from the PnP framework combined with the EDN-based denoiser, which jointly captures structured sparsity and spreading effects and generalizes better than purely data-driven networks.} \par  
{In addition, we evaluate the robustness of the algorithm under different dispersive OTFS channel conditions\footnote{{Leveraging the inherent duality between the Doppler and delay domains, the analytical principles governing fractional Doppler are directly applicable to fractional delay. Therefore, rather than developing a redundant analytical model, we synthesize the effects of fractional delay numerically to evaluate its impact.}}, as shown in Fig.~\ref{Fractional_Channel}. When only fractional Doppler is present, the proposed method yields lower estimation error since the leakage occurs mainly along the Doppler dimension and the dominant channel taps remain distinguishable. When both fractional delay and fractional Doppler exist, the channel sparsity is further reduced and interference spreads over two dimensions, leading to additional degradation. By contrast, variations in the number of paths have only a limited influence, as the overall channel structure is preserved and PnP-CE remains robust against moderate changes in path diversity. This is because the implicit prior learned by the denoiser captures local spreading patterns more effectively than deterministic priors, allowing the algorithm to retain robustness. } \par 
{Furthermore, we compare the proposed PnP-CE with sparse channel estimation baselines under the fractional delay and Doppler scenario, as shown in Fig.~\ref{different_priors}. The ADMM estimator without any prior produces the poorest NMSE, since the solution is unconstrained. When an $\ell_{1}$ prior is introduced, the performance becomes highly sensitive to the choice of regularization strength. We present three configurations, where the low threshold results in insufficient suppression of noise, the high threshold discards valid channel components, and the adaptive threshold adjusts the parameter according to noise statistics and system dimensions, thereby alleviating part of this sensitivity. Although the adaptive approach improves robustness, its gains remain modest, as the $\ell_{1}$ prior cannot adequately characterize the energy spreading caused by fractional delay and Doppler. In comparison, the EM-SBL algorithm with a hierarchical prior provides more stable results by exploiting a stronger statistical model, while the proposed PnP-CE achieves the best performance owing to its learned denoiser, which acts as a flexible implicit prior and adapts to complex channel structures.}

\begin{figure}[t]
    \centering
    \begin{subfigure}{0.45\linewidth}
      \centering
      \includegraphics[width=\linewidth]{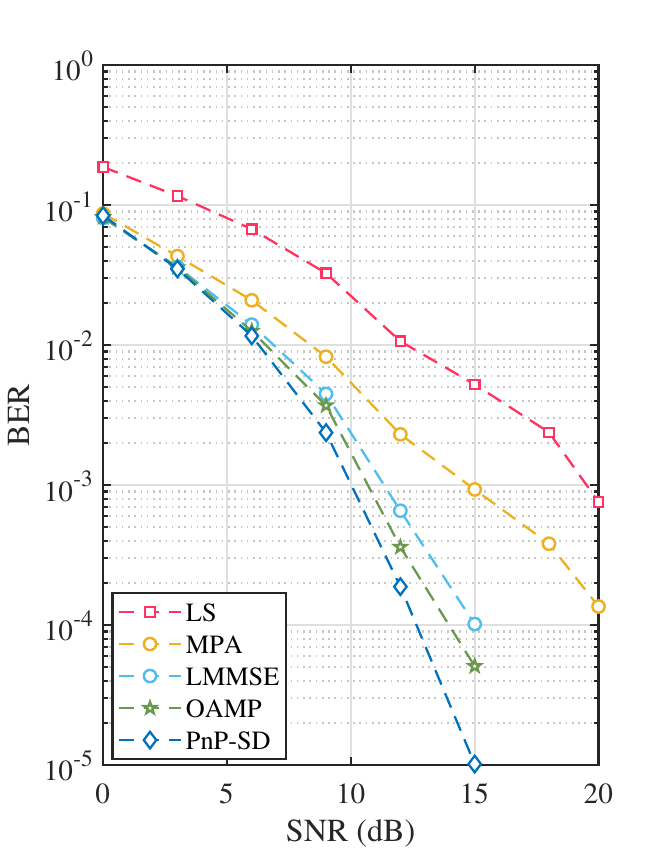}
      \caption{{Integer Doppler.}}
      \label{Baselines_L1}
    \end{subfigure}
    \hfill
    \begin{subfigure}{0.45\linewidth}
      \centering
      \includegraphics[width=\linewidth]{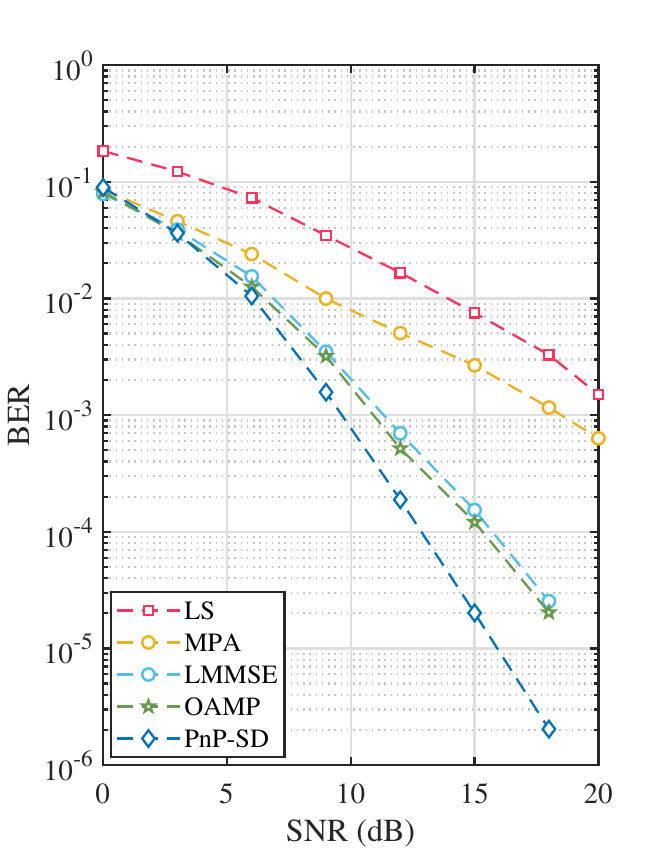}
      \caption{{Fractional Doppler.}}
      \label{Baselines_R1}
    \end{subfigure}
    \caption{BER performance comparison with baselines for symbol detection.}
    \label{Baselines1}
\end{figure}
\begin{figure}[t]
    \centering
    \begin{subfigure}{0.45\linewidth}
      \centering
      \includegraphics[width=\linewidth]{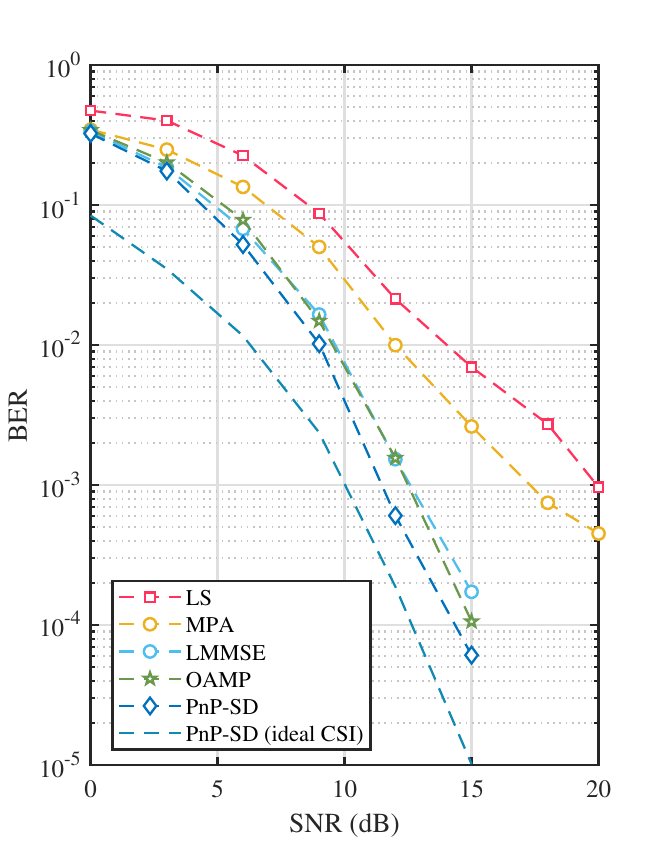}
      \caption{{BER comparison with slightly imperfect CSI.}}
      \label{Baselines_L2}
    \end{subfigure}
    \hfill
    \begin{subfigure}{0.45\linewidth}
      \centering
      \includegraphics[width=\linewidth]{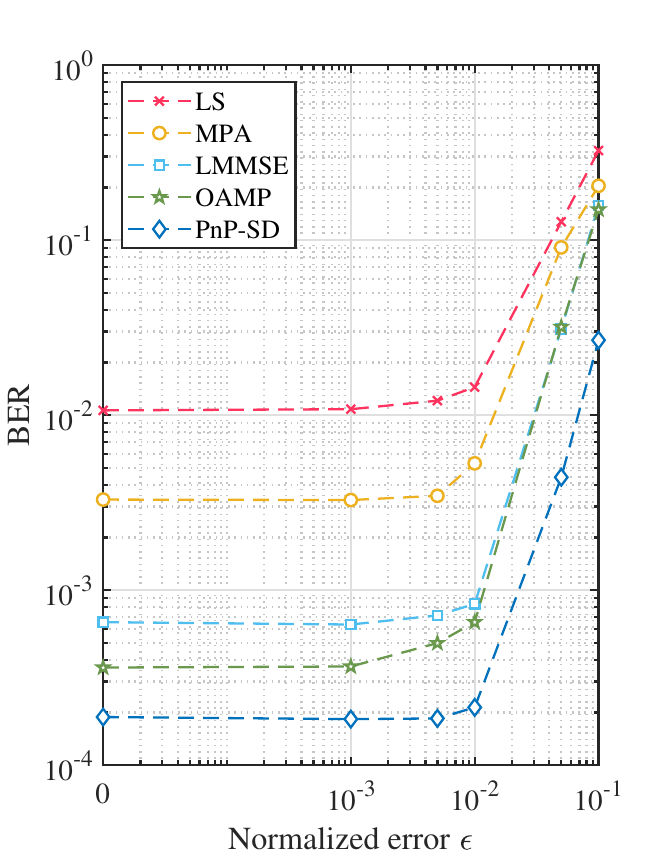}
      \caption{{BER comparison with normalized error.}}
      \label{Baselines_R2}
    \end{subfigure}
    \caption{BER comparison with non-ideal channel.}
    \label{Baselines2}
\end{figure}

\subsection{Results for Symbol Detection}
{In this section, we first investigate the reliability of the PnP-SD algorithm. In particular, we evaluate the bit error rate (BER) performance under different received SNRs for various symbol detection algorithms,  including LS, LMMSE, MPA \cite{li2023otfs}, and orthogonal approximate message passing (OAMP) \cite{haifeng2022cross}.} Fig. \ref{Baselines_L1} presents the curves of BER versus SNR with the integer delay and Doppler in the OTFS effective channel. It can be observed that the LS method has the worst performance since no prior noise statistics are utilized in the detection. We can also observe that the MPA outperforms the LS estimator. This is because the MPA can effectively incorporate prior knowledge, such as sparsity or the statistical properties of the channel. The LMMSE algorithm achieves better performance as it treats the transmitted symbol prior as a Gaussian distribution. {The OAMP detector further improves over LMMSE by iteratively refining the estimates with message passing and exploiting the structured residual, but its performance remains below PnP-SD due to limited adaptability to non-Gaussian interference.} In contrast, the proposed PnP-SD algorithm exhibits the best performance as the PnP-SD employs a trained denoiser to adaptively mitigate noise interference. As shown in Fig.~\ref{Baselines_R1}, the performance of all algorithms degrades in the presence of fractional Doppler. The MPA algorithm is the most affected, since its detection relies heavily on strict channel sparsity, which no longer holds when energy spreads across Doppler bins. {The OAMP method also suffers performance loss, as its iterative refinement assumes structured residuals that become less accurate under fractional Doppler. In contrast, the PnP-SD exhibits greater robustness, maintaining stable performance even under reduced sparsity.}

{Furthermore, with the normalized error $\epsilon$ classified as slightly imperfect when $\epsilon \le 10^{-2}$, we investigate the impact of the channel estimation error on the BER performance of different algorithms, as shown in Fig. \ref{Baselines_L2}.} Among the evaluated algorithms, the PnP-SD algorithm consistently exhibits superior performance, achieving the lowest BER across the full SNR range. These results are expected since our proposed method with the embedded neural network can learn the features of the symbols and interference to guarantee the BER performance. {For reference, we also include the performance of the proposed receiver under perfect CSI, which highlights the SNR degradation relative to the ideal case.} {Fig. \ref{Baselines_R2} illustrates BER performance at a fixed SNR of 12 dB as the normalized CSI estimation error increases. At small error levels, all algorithms maintain relatively stable performance, with PnP-SD achieving the lowest BER. As the error grows, LS and MPA exhibit the most rapid degradation, while LMMSE achieves better robustness but still suffers noticeable performance loss. {The performance of the OAMP detector degrades more rapidly beyond $10^{-3}$, and the degradation eventually approaches that of LMMSE. This is because OAMP relies on accurate residual updates and approximate Gaussianity assumptions, which break down more severely under large CSI errors.} In contrast, the proposed PnP-SD algorithm demonstrates the highest resilience, maintaining significantly lower BER across a wide range of error levels.}\par
\begin{figure}[tb]
    \centering
    \includegraphics[width=0.85\linewidth]{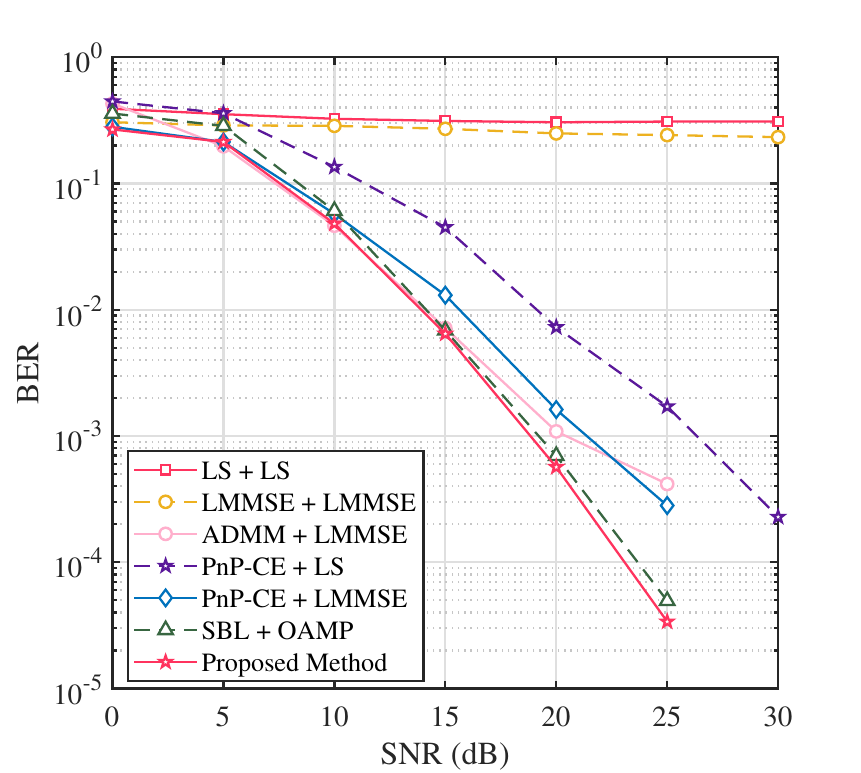}
    \caption{{BER performance for sequential channel estimation and symbol detection.}}
    \label{Joint_CESD}
\end{figure}

{We investigate the BER performance of sequential channel estimation and symbol detection, where channel estimation is first performed and the resulting CSI is then used for symbol detection.} {As illustrated in Fig. \ref{Joint_CESD}, we evaluate our proposed method against four baselines: (1) LS + LS, (2) LMMSE + LMMSE, (3) ADMM + LMMSE, (4) PnP-CE + LS, (5) PnP-CE + LMMSE, (6) EM-SBL+OAMP, and (7) our proposed method (PnP-CE + PnP-SD).} It can be observed that the conventional LS + LS and LMMSE + LMMSE approaches demonstrate the poorest performance, particularly at low SNR, since their limited channel estimation accuracy directly propagates to symbol detection. The ADMM + LMMSE method provides a noticeable improvement, especially in the low-SNR regime, because the iterative optimization in ADMM produces more reliable channel estimates. {The PnP-CE + LS baseline shows consistent gains over LS + LS, demonstrating that the PnP-based channel estimate offers tangible benefits even when paired with a simple detector, although its performance is still inferior to detection schemes that employ MMSE or PnP-SD.} The PnP-CE + LMMSE approach further improves detection accuracy by leveraging the enhanced channel estimates, outperforming both ADMM + LMMSE and conventional methods, particularly at higher SNRs. {The SBL + OAMP method performs worse at low SNR due to the sensitivity of SBL estimation in noisy conditions, but its performance becomes stable and surpasses other traditional baselines at medium-to-high SNR owing to the robustness of SBL estimation and the effectiveness of OAMP detection.} Finally, the proposed method (PnP-CE + PnP-SD) achieves the best BER performance across the entire SNR range, as the use of PnP-based estimation and detection enables accurate CSI recovery and robust interference mitigation.\par

\begin{figure}[t]
    \centering
    \begin{subfigure}{0.45\linewidth}
      \centering
      \includegraphics[width=\linewidth]{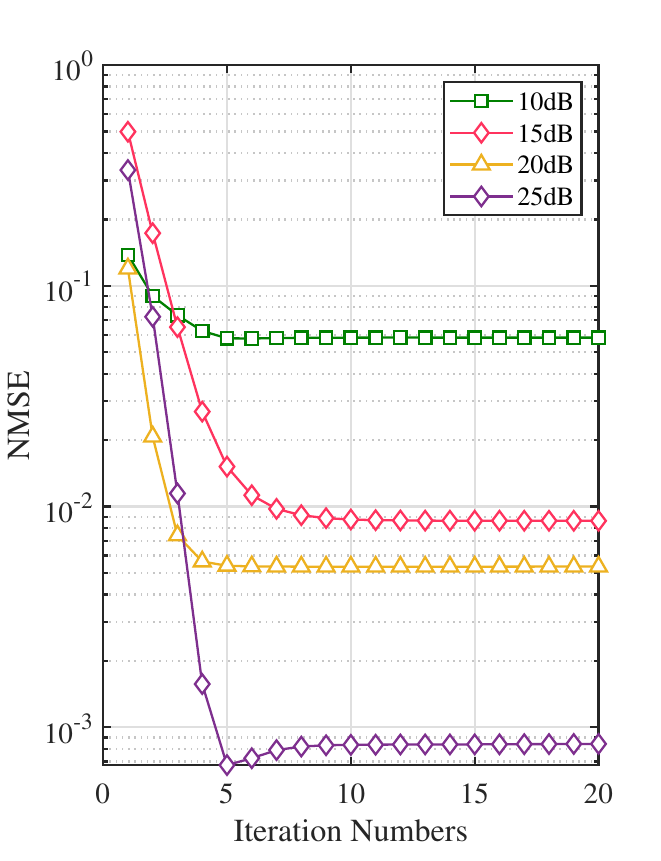}
      \caption{{The convergence of the PnP-CE method.}}
      \label{Cver1_Channel}
    \end{subfigure}
    \hfill
    \begin{subfigure}{0.45\linewidth}
      \centering
      \includegraphics[width=\linewidth]{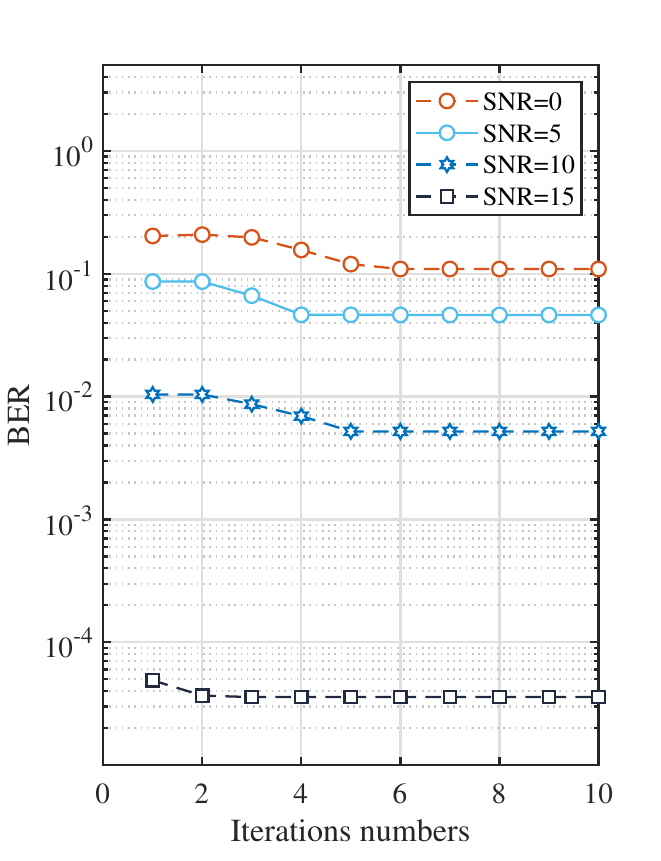}
      \caption{{The convergence of the PnP-SD method.}}
      \label{Cver1_Symbol}
    \end{subfigure}
    \caption{Empirical evaluation of convergence for the proposed PnP framework.}
    \label{Converagence}
\end{figure}

\vspace{-0.3cm}
\subsection{Convergence}
Recent theoretical studies have extensively analyzed the convergence properties of deep PnP prior algorithms. In \cite{sreehari2016plug}, sufficient conditions were established by requiring the denoising operator to behave as a proximal mapping, and subsequent work further provided guarantees under assumptions such as boundedness and continuity of the denoiser \cite{mukherjee2023learned}. Empirical studies have also confirmed the practical effectiveness of DL-based denoisers in the PnP framework \cite{ryu2019plug}. In the following, we empirically evaluate the convergence of the PnP-CE and PnP-SD algorithms. As shown in Fig.~\ref{Cver1_Channel}, the NMSE of PnP-CE decreases sharply within the first few iterations, demonstrating fast error reduction. Similarly, Fig.~\ref{Cver1_Symbol} presents the BER convergence of PnP-SD under different SNRs for a representative channel realization, where the error rate drops rapidly at the beginning and then stabilizes. These results indicate that both algorithms converge reliably within only a few iterations. It is noteworthy that the proposed framework implicitly exploits the DL-based denoiser as a regularizer, and the observed convergence behavior shows that such denoisers introduce implicit regularization effects that help stabilize the iterative process.

\section{Conclusions}  

In this paper, we investigated DL-based channel estimation and symbol detection for OTFS by developing a unified PnP framework. The optimization objective was decomposed into a prior-related term and a data-fidelity term. The prior-related term was treated as a denoising task, effectively handled using deep learning. To this end, we incorporated a pre-trained lightweight EDN into the framework, yielding the PnP-CE method. For symbol detection, a trained MLP-based denoiser was similarly integrated. Moreover, we simplified the model-based step by leveraging the time-domain measurement matrix, improving robustness and reducing complexity. The proposed hybrid framework preserves the adaptability of traditional optimization while harnessing the representational strength of deep learning, enhancing OTFS performance under challenging wireless conditions. {This work has focused on the single-antenna case, and the PnP framework could be extended to MIMO-OTFS systems through its linear inverse formulation, which will be considered in future work.}

\bibliographystyle{IEEEtran}
\bibliography{main}

@article{shen2019channel,
  title={Channel estimation for orthogonal time frequency space ({OTFS}) massive {MIMO}},
  author={Shen, Wenqian and Dai, Linglong and An, Jianping and Fan, Pingzhi and Heath, Robert W},
  journal={IEEE Trans. Signal Process.},
  volume={67},
  number={16},
  pages={4204--4217},
  year={2019},
  publisher={IEEE}
}

@ARTICLE{Zhao2020SBL,
  author={Zhao, Lei and Gao, Wen-Jing and Guo, Wenbin},
  journal={IEEE Commun. Lett.}, 
  title={Sparse Bayesian Learning of {Delay-Doppler} Channel for {OTFS} System}, 
  year={2020},
  volume={24},
  number={12},
  pages={2766-2769},
  doi={10.1109/LCOMM.2020.3021120}}

@article{yuan2020simple,
  title={A simple variational Bayes detector for orthogonal time frequency space ({OTFS}) modulation},
  author={Yuan, Weijie and Wei, Zhiqiang and Yuan, Jinhong and Ng, Derrick Wing Kwan},
  journal={IEEE Trans. Veh. Technol.},
  volume={69},
  number={7},
  month={Jul.},
  pages={7976--7980},
  year={2020},
  publisher={IEEE}
}

@article{raviteja2019embedded,
  title={Embedded pilot-aided channel estimation for {OTFS} in delay--Doppler channels},
  author={Raviteja, Patchava and Phan, Khoa T and Hong, Yi},
  journal={IEEE Trans. Veh. Technol.},
  volume={68},
  number={5},
  pages={4906--4917},
  year={2019},
  month = {May.},
  publisher={IEEE}
}

@ARTICLE{qing2023viterbi,
  author={Gong, Yi and Li, Qingyu and Meng, Fanke and Li, Xinru and Xu, Zhan},
  journal={IEEE Commun. Lett.}, 
  title={ViterbiNet-Based Signal Detection for {OTFS} System}, 
  year={2023},
  month={Mar.},
  volume={27},
  number={3},
  pages={881-885},
  doi={10.1109/LCOMM.2023.3237719}}

@ARTICLE{xiaoqi2024sparse,
  author={Zhang, Xiaoqi and Liu, Chang and Yuan, Weijie and Zhang, J. Andrew and Ng, Derrick Wing Kwan},
  journal={IEEE Trans. Veh. Technol.}, 
  title={Sparse Prior-Guided Deep Learning for {OTFS} Channel Estimation}, 
  year={2024},
  volume={},
  number={},
  pages={1-6},
  doi={10.1109/TVT.2024.3450012}}

@article{wang2022joint,
  title={Joint Bayesian channel estimation and data detection for {OTFS} systems in {LEO} satellite communications},
  author={Wang, Xueyang and Shen, Wenqian and Xing, Chengwen and An, Jianping and Hanzo, Lajos},
  journal={IEEE Trans. Commun.},
  volume={70},
  number={7},
  pages={4386--4399},
  year={2022},
  month={Jul.},
  publisher={IEEE}
}

@article{zhang2021plug,
  title={Plug-and-play image restoration with deep denoiser prior},
  author={Zhang, Kai and Li, Yawei and Zuo, Wangmeng and Zhang, Lei and Van Gool, Luc and Timofte, Radu},
  journal={IEEE Trans. Pattern Anal. Mach. Intell.},
  volume={44},
  number={10},
  pages={6360--6376},
  year={2021},
  month={Oct.},
  publisher={IEEE}
}

@article{lai2022deep,
  title={Deep plug-and-play prior for hyperspectral image restoration},
  author={Lai, Zeqiang and Wei, Kaixuan and Fu, Ying},
  journal={Neurocomputing},
  volume={481},
  pages={281--293},
  year={2022},
  publisher={Elsevier}
}

@ARTICLE{wei2021waveform,
  author={Wei, Zhiqiang and Yuan, Weijie and Li, Shuangyang and Yuan, Jinhong and Bharatula, Ganesh and Hadani, Ronny and Hanzo, Lajos},
  journal={IEEE Wireless Commun.}, 
  title={Orthogonal Time-Frequency Space Modulation: A Promising Next-Generation Waveform}, 
  year={2021},
  month={Apr.},
  volume={28},
  number={4},
  pages={136-144},
  doi={10.1109/MWC.001.2000408}}

@article{wei2021transmitter,
  title={Transmitter and receiver window designs for orthogonal time-frequency space modulation},
  author={Wei, Zhiqiang and Yuan, Weijie and Li, Shuangyang and Yuan, Jinhong and Ng, Derrick Wing Kwan},
  journal={IEEE Trans. Commun.},
  volume={69},
  number={4},
  pages={2207--2223},
  year={2021},
month={Apr.},
  publisher={IEEE}
}

@ARTICLE{Liu2022RIS,
  author={Liu, Chang and Liu, Xuemeng and Ng, Derrick Wing Kwan and Yuan, Jinhong},
  journal={IEEE Trans. Wireless Commun.}, 
  title={Deep Residual Learning for Channel Estimation in Intelligent Reflecting Surface-Assisted Multi-User Communications}, 
  year={2022},
  month={Feb.},
  volume={21},
  number={2},
  pages={898-912},
  doi={10.1109/TWC.2021.3100148}
}

@ARTICLE{Fei2021Message,
  author={Liu, Fei and Yuan, Zhengdao and Guo, Qinghua and Wang, Zhongyong and Sun, Peng},
  journal={IEEE Trans. Wireless Commun.}, 
  title={Message Passing-Based Structured Sparse Signal Recovery for Estimation of {OTFS} Channels With Fractional Doppler Shifts}, 
  year={2021},
  month={Dec.},
  volume={20},
  number={12},
  pages={7773-7785},
  doi={10.1109/TWC.2021.3087501}}

@article{zhang2024wireless,
  title={Wireless Communications in Doubly Selective Channels with Domain Adaptivity},
  author={Zhang, J Andrew and Zhang, Hongyang and Wu, Kai and Huang, Xiaojing and Yuan, Jinhong and Guo, Y Jay},
  journal={Accepted for IEEE Communications Magazine},
  year={2024}
}

@inproceedings{li2022uamp,
  title={UAMP-based channel estimation for {OTFS} in the presence of the fractional doppler with HMM prior},
  author={Li, Zhongjie and Yuan, Weijie and Zhou, Lin},
  booktitle={2022 IEEE/CIC Int. Conf. Commun. China (ICCC Workshops)},
  pages={304--308},
  year={2022},
  organization={IEEE}
}

@inproceedings{ryu2019plug,
  title={Plug-and-play methods provably converge with properly trained denoisers},
  author={Ryu, Ernest and Liu, Jialin and Wang, Sicheng and Chen, Xiaohan and Wang, Zhangyang and Yin, Wotao},
  booktitle={Int. Conf. Mach. Learning},
  pages={5546--5557},
  year={2019},
  organization={PMLR}
}

@ARTICLE{Wan2024Multitask,
  author={Wan, Weixiao and Chen, Wei and Wang, Shiyue and Li, Geoffrey Ye and Ai, Bo},
  journal={IEEE Trans. Commun.}, 
  title={Deep Plug-and-Play Prior for Multitask Channel Reconstruction in Massive {MIMO} Systems}, 
  year={2024},
  month={Jul.},
  volume={72},
  number={7},
  pages={4149-4162},
  doi={10.1109/TCOMM.2024.3369702}}

@article{sreehari2016plug,
  title={Plug-and-play priors for bright field electron tomography and sparse interpolation},
  author={Sreehari, Suhas and Venkatakrishnan, S Venkat and Wohlberg, Brendt and Buzzard, Gregery T and Drummy, Lawrence F and Simmons, Jeffrey P and Bouman, Charles A},
  journal={IEEE Trans. Comput. Imag.},
  volume={2},
  number={4},
  pages={408--423},
  year={2016},
  month={Apr.},
  publisher={IEEE}
}

@ARTICLE{wei2022offgrid,
  author={Wei, Zhiqiang and Yuan, Weijie and Li, Shuangyang and Yuan, Jinhong and Ng, Derrick Wing Kwan},
  journal={IEEE Trans. Wireless Commun.}, 
  title={Off-Grid Channel Estimation With Sparse Bayesian Learning for OTFS Systems}, 
  year={2022},
  volume={21},
  number={9},
  pages={7407-7426},
  doi={10.1109/TWC.2022.3158616}}

@article{mukherjee2023learned,
  title={Learned reconstruction methods with convergence guarantees: A survey of concepts and applications},
  author={Mukherjee, Subhadip and Hauptmann, Andreas and {\"O}ktem, Ozan and Pereyra, Marcelo and Sch{\"o}nlieb, Carola-Bibiane},
  journal={IEEE Signal Process. Mag.},
  volume={40},
  number={1},
  pages={164--182},
  year={2023},
  month={Jan.},
  publisher={IEEE}
}

@inproceedings{he2016deep,
  title={Deep residual learning for image recognition},
  author={He, Kaiming and Zhang, Xiangyu and Ren, Shaoqing and Sun, Jian},
  booktitle={Proceedings of the IEEE Conf. Comput. Vis. Pattern Recog.},
  pages={770--778},
  year={2016}
}

@inproceedings{yoo2004capacity,
  title={Capacity of fading {MIMO} channels with channel estimation error},
  author={Yoo, Taesang and Goldsmith, Andrea},
  booktitle={2004 IEEE Int. Conf. Commun. (IEEE Cat. No. 04CH37577)},
  volume={2},
  pages={808--813},
  year={2004},
  organization={IEEE}
}

@article{liu2023predictive,
  title={Predictive precoder design for {OTFS}-enabled {URLLC}: A deep learning approach},
  author={Liu, Chang and Li, Shuangyang and Yuan, Weijie and Liu, Xuemeng and Ng, Derrick Wing Kwan},
  journal={IEEE J. Sel. Areas Commun.},
  volume={41},
  number={7},
  pages={2245--2260},
  year={2023},
  month={Jul.},
  publisher={IEEE}
}

@article{shen2022random,
  title={Random access with massive MIMO-OTFS in {LEO} satellite communications},
  author={Shen, Boxiao and Wu, Yongpeng and An, Jianping and Xing, Chengwen and Zhao, Lian and Zhang, Wenjun},
  journal={IEEE J. Sel. Areas Commun.},
  volume={40},
  number={10},
  pages={2865--2881},
  year={2022},
  month={Oct.},
  publisher={IEEE}
}

@INPROCEEDINGS{haifeng2022cross,
  author={Wen, Haifeng and Yuan, Weijie and Wu, Nan and Wen, Jinming},
  booktitle={2022 Int. Sympos. Wireless Commun. Syst. (ISWCS)}, 
  title={A Low-Complexity Cross-Domain {OAMP} Detector for {OTFS}}, 
  year={2022},
  volume={},
  number={},
  pages={1-6},
  doi={10.1109/ISWCS56560.2022.9940366}}

@INPROCEEDINGS{zhongjie2022uamp,
  author={Li, Zhongjie and Yuan, Weijie and Zhou, Lin},
  booktitle={2022 IEEE/CIC Int. Conf. Commun. China (ICCC Workshops)}, 
  title={UAMP-Based Channel Estimation for {OTFS} in the Presence of the Fractional Doppler with {HMM} Prior}, 
  year={2022},
  volume={},
  number={},
  pages={304-308},
  doi={10.1109/ICCCWorkshops55477.2022.9896709}}

@article{ye2017power,
  title={Power of deep learning for channel estimation and signal detection in {OFDM} systems},
  author={Ye, Hao and Li, Geoffrey Ye and Juang, Biing-Hwang},
  journal={IEEE Wireless Commun. Lett.},
  volume={7},
  number={1},
  month={Jan.},
  pages={114--117},
  year={2017},
  publisher={IEEE}
}

@article{li2023otfs,
  title={{OTFS} detection based on gaussian mixture distribution: A generalized message passing approach},
  author={Li, Xiang and Yuan, Weijie},
  journal={IEEE Commun. Lett.},
  volume={28},
  number={1},
  pages={178--182},
  year={2023},
  month={Jan.},
  publisher={IEEE}
}

@article{yuan2018iterative,
  title={Iterative receivers for downlink {MIMO-SCMA}: Message passing and distributed cooperative detection},
  author={Yuan, Weijie and Wu, Nan and Guo, Qinghua and Li, Yonghui and Xing, Chengwen and Kuang, Jingming},
  journal={IEEE Trans. Wireless Commun.},
  volume={17},
  number={5},
  pages={3444--3458},
  year={2018},
  month={May.},
  publisher={IEEE}
}

@article{ni2021uplink,
  title={Uplink sensing in perceptive mobile networks with asynchronous transceivers},
  author={Ni, Zhitong and Zhang, J Andrew and Huang, Xiaojing and Yang, Kai and Yuan, Jinhong},
  journal={IEEE Trans. Signal Process.},
  volume={69},
  pages={1287--1300},
  year={2021},
  publisher={IEEE}
}

@article{nie2024uplink,
  title={Uplink Multi-User OTFS: Transmitter Design based on Statistical Channel Information},
  author={Nie, Mingcheng and Li, Shuangyang and Mishra, Deepak and Yuan, Jinhong and Ng, Derrick Wing Kwan},
  journal={IEEE Trans. Commun.},
  year={2024},
  publisher={IEEE}
}

@article{lecun2015deep,
  title={Deep learning},
  author={LeCun, Yann and Bengio, Yoshua and Hinton, Geoffrey},
  journal={nature},
  volume={521},
  number={7553},
  pages={436--444},
  year={2015},
  publisher={Nature Publishing Group UK London}
}

@book{bauschke2017correction,
  title={Correction to: convex analysis and monotone operator theory in Hilbert spaces},
  author={Bauschke, Heinz H and Combettes, Patrick L and Bauschke, Heinz H and Combettes, Patrick L},
  year={2017},
  publisher={Springer}
}

\end{document}